\documentclass[twocolumn,preprintnumbers]{revtex4}

\usepackage{graphicx}
\usepackage{epsf}
\usepackage{amsmath,amssymb}
\usepackage{color}
\newcommand{\bvec}{\boldsymbol}

\newcommand{\Be}{^{10}\textrm{Be}}
\newcommand{\C}{^{10}\textrm{C}}
\newcommand{\Mg}{^{26}\textrm{Mg}}

\usepackage{ulem}

\begin{document}
\preprint{KUNS 2812, NITEP 67}
\title{Neutron dominance in excited states of $^{26}$Mg and $^{10}$Be probed 
by proton and alpha inelastic scattering}

\author{Yoshiko Kanada-En'yo$^{1}$}
\author{Yuki Shikata$^{1}$}
\author{Yohei~Chiba$^{2,3}$}
\author{Kazuyuki Ogata$^{2,3,4}$} 
\affiliation{$^{1}$Department of Physics, Kyoto University, Kyoto 606-8502, Japan}
\affiliation{$^{2}$Department of Physics, Osaka City University, Osaka 558-8585, Japan}
\affiliation{$^{3}$Research Center for Nuclear Physics (RCNP), Osaka
University, Ibaraki 567-0047, Japan}
\affiliation{$^{4}$
Nambu Yoichiro Institute of Theoretical and Experimental Physics (NITEP),
   Osaka City University, Osaka 558-8585, Japan}

\begin{abstract}
Isospin characters of nuclear excitations in $^{26}$Mg and $^{10}$Be are investigated
via proton~($p$) and alpha~($\alpha$) inelastic scattering.
A structure model of antisymmetrized molecular dynamics (AMD) is applied to calculate the
ground and excited states of $^{26}$Mg and $^{10}$Be.
The calculation describes the isoscalar feature of the ground-band $2^+_1$($K^\pi=0^+_1$) excitation and predicts the neutron dominance of the side-band $2^+_2$($K^\pi=2^+$) excitation 
in $^{26}$Mg and $^{10}$Be. 
The $p$ and $\alpha$ inelastic scattering off $^{26}$Mg and $^{10}$Be is calculated by 
microscopic coupled-channel (MCC) calculations with a $g$-matrix folding approach 
by using the matter and transition densities of the target nuclei calculated with AMD.
The calculation reasonably reproduces 
the observed $0^+_1$, $2^+_1$, and $2^+_2$ cross sections of $^{26}$Mg+$p$ scattering 
at incident energies 
$E_p=24$ and 40~MeV and of $^{26}$Mg+$\alpha$ scattering at $E_\alpha=104$ and 120~MeV.
For $^{10}$Be+$p$ and  $^{10}$Be+$\alpha$ scattering, inelastic cross sections to the excited states in 
the $K^\pi=0^+_1$ ground-, $K^\pi=2^+$ side-, $K^\pi=0^+_2$ cluster-, and $K^\pi=1^-$ cluster-bands 
are investigated.
The isospin characters of excitations are investigated via inelastic scattering processes 
by comparison of the production rates in the $^{10}$Be+$p$, $^{10}$Be+$\alpha$, and  $^{10}$C+$p$ reactions. 
The result predicts that the  $2^+_2$ state is selectively produced by the $^{10}$Be+$p$ reaction 
because of the neutron dominance in the  $2^+_2$ excitation as in the case of the $^{26}$Mg+$p$ 
scattering to the $2^+_2$ state,  
whereas its production 
is significantly suppressed in the $^{10}$C+$p$ reaction. 
\end{abstract}

\maketitle

\section{Introduction}

Isospin characters of nuclear excitations in $Z\ne N$ nuclei
have been attracting great interests. 
To discuss the difference between neutron and proton components 
in nuclear deformations and excitations,
the neutron and proton transition matrix elements, $M_n$ and $M_p$,
have been extensively investigated 
by experimental works with mirror analysis of electric transitions 
and hadron inelastic scattering with $\alpha$, $p$, and $\pi^-/\pi^+$ 
as well as electron inelastic scattering. The ratio $M_n/M_p$ 
has been discussed with the isoscalar and isovector components of $2^+$ excitations 
for various stable nuclei~\cite{Bernstein:1977wtr,Bernstein:1979zza,Bernstein:1981fp,Brown:1980zzd,Brown:1982zz}.
The simple relation $M_n/M_p=N/Z$ is naively expected for a uniform rigid rotor model, 
while $M_n/M_p=1$ should be satisfied if only a $Z=N$ core part contributes to the excitation. 
In the analysis of the $M_n/M_p$ ratio, it has been reported that 
$M_n/M_p$ systematically exceeds $N/Z$ in proton closed-shell nuclei. In particular, 
an extremely large value of the ratio was found in $^{18}$O, which expresses remarkable 
neutron dominance of the $2^+_1$ excitation. 
In the opposite case, $M_n/M_p < N/Z$ of proton dominance was obtained in neutron closed-shell nuclei.

For $^{26}$Mg, the $M_n/M_p$ ratio has been investigated for various excited states by means of 
life-time measurements of mirror transitions~\cite{Alexander:1982sph}, and 
$\pi^-/\pi^+$, 
$p$, and $\alpha$~\cite{Wiedner:1980cea,Alons:1981rrm,Zwieglinski:1983zz,VanDerBorg:1981qiu} inelastic scattering. 
In those analyses, the strong state dependency of isospin characters has been found in 
the first and second $2^+$ states.
The ratio $M_n/M_p= 0.7$--1 was obtained for the $0^+_1\to 2^+_1$ transition, whereas
$M_n/M_p= 1.2$--4 was estimated for the $0^+_1\to 2^+_2$ transition. The former indicates an approximately 
isoscalar nature of the $2^+_1$ excitation, while the latter shows predominant neutron component
of the $2^+_2$ excitation. However, there remains significant uncertainty in the neutron component of the $0^+_1\to 2^+_2$ transition. 

The isospin characters of nuclear excitations are hot issues also in the physics of unstable nuclei. 
The neutron dominance in the $2^+_1$ state has been
suggested in neutron-rich nuclei such as $^{12}$Be and 
$^{16}$C~\cite{Kanada-Enyo:1996zsp,Iwasaki:2000gh,Kanada-Enyo:2004tao,Kanada-Enyo:2004ere,Sagawa:2004ut,Takashina:2005bs,Ong:2006rm,Burvenich:2008zz,Takashina:2008zza,Elekes:2008zz,Wiedeking:2008zzb, Yao:2011zza, Forssen:2011dr}.
The proton component can be determined from $B(E2)$ measured by $\gamma$ decays.
For the neutron component, such tools as mirror analysis and $\pi^-/\pi^+$ scattering are practically difficult for neutron-rich nuclei.  
Instead, $p$ inelastic scattering experiments in the inverse kinematics
have been intensively performed to probe the neutron component
and supported the neutron dominance in the $2^+_1$ state of $^{12}$Be and $^{16}$C. 
Very recently, Furuno {\it et al.} have achieved an $\alpha$ inelastic 
scattering experiment off $^{10}\textrm{C}$ 
in the inverse kinematics and discussed the isospin characters of 
the $2^+_1$ excitation~\cite{Furuno:2019lyp}.

Our aim in this paper is to investigate isospin characters of the $2^+_1$ and $2^+_2$ excitations in $\Mg$ and 
$\Be$ with microscopic coupled-channel (MCC) calculations of $p$ and $\alpha$ scattering. 
We also aim to predict inelastic cross sections to cluster excitations of $\Be$. 
Structures of the ground and excited states of $\Be$ have been studied with many theoretical models, 
and described well by the cluster structure of $2\alpha+nn$
(see Refs.~\cite{Oertzen-rev,KanadaEn'yo:2012bj,Ito2014-rev} and references therein). 
In Ref.~\cite{Kanada-Enyo:2011plo}, one of the authors (Y. K-E.) 
has discussed the $2^+_1$ and $2^+_2$ excitations of  $\Be$  with the isovector triaxiality,
and predicted the neutron dominance in the $2^+_2$ excitation similarly to that of $\Mg$. 

In the present MCC calculations, the nucleon-nucleus potentials
are microscopically derived by folding the Melbourne $g$-matrix $NN$ interaction with 
diagonal and transition densities of target nuclei, which are obtained from microscopic structure models.
The $\alpha$-nucleus potentials are obtained by folding the 
nucleon-nucleus potentials with an  $\alpha$ density. 
The MCC approach with the Melbourne $g$-matrix $NN$ interaction 
has successfully described the observed cross sections of 
$p$ and $\alpha$ elastic and inelastic scattering off 
various nuclei
at $p$ energies from 40~MeV to 300~MeV and $\alpha$ energies from 100~MeV to 
400~MeV~\cite{Amos:2000,Karataglidis:2007yj,Minomo:2009ds,Toyokawa:2013uua,Minomo:2017hjl,Egashira:2014zda,Minomo:2016hgc}. 
In our recent works~\cite{Kanada-Enyo:2019prr,Kanada-Enyo:2019qbp,Kanada-Enyo:2019uvg,Kanada-Enyo:2020zpl},  we have applied the MCC calculations by using
matter and transition densities of target nuclei 
calculated by a structure model of antisymmetrized molecular dynamics (AMD)~\cite{KanadaEnyo:1995tb,KanadaEn'yo:1998rf,Kanada-Enyo:1999bsw,KanadaEn'yo:2012bj} and 
investigated transition properties of low-lying states of various stable and unstable 
nuclei via $p$ and $\alpha$ inelastic scattering. 
One of the advantages of this approach is that one can discuss inelastic processes of 
different hadronic probes, $p$ and $\alpha$, in a unified treatment of a microscopic description.
Another advantage is that there is no phenomenological parameter in the reaction part.
Since one can obtain cross sections at given energies for given structure inputs with no ambiguity, 
it can test the validity of the structure inputs via $p$ and $\alpha$ cross sections 
straightforwardly. 

In this paper, we apply the MCC approach to $p$ and $\alpha$ scattering off $\Mg$ and 
$\Be$ using the AMD densities of the target nuclei, and 
investigate isospin characters of inelastic transitions of  $\Mg$ and $\Be$.
Particular attention is paid on transition features of the 
ground-band $2^+_1$ state and the side-band $2^+_2$ state.
We also give theoretical a prediction of inelastic cross sections to cluster states of $A=10$ nuclei
of the $\Be+p$, $\Be+\alpha$, and $\C+p$ reactions. 

The paper is organized as follows. 
The next section briefly describes the MCC approach for the reaction calculations of $p$ and $\alpha$ scattering and the AMD framework for structure calculations of $\Mg$ and $\Be$.
Structure properties of $\Mg$ and $\Be$ are described in Sec.~\ref{sec:results1}, 
and transition properties and $p$ and $\alpha$ scattering are discussed 
in Sec.~\ref{sec:results2}.
Finally, the paper is summarized in Sec.~\ref{sec:summary}.

\section{Method} \label{sec:method} 

The reaction calculations of $p$ and $\alpha$ scattering are performed
with the MCC approach as 
done in Refs.~\cite{Kanada-Enyo:2019prr,Kanada-Enyo:2019qbp,Kanada-Enyo:2019uvg}.
The diagonal and coupling potentials for the nucleon-nucleus system are microscopically calculated
by folding the Melbourne $g$-matrix $NN$ interaction~\cite{Amos:2000} with densities of the target nucleus calculated by AMD. 
The $\alpha$-nucleus potentials are obtained in an extended nucleon-nucleus
folding model~\cite{Egashira:2014zda} by folding the nucleon-nucleus potentials 
with an $\alpha$ density given by a one-range Gaussian form.
In the present reaction calculations, 
the spin-orbit term of the potentials is not taken into account to avoid complexity as in Refs.~\cite{Kanada-Enyo:2019uvg,Kanada-Enyo:2020zpl}.
It should be stressed again that there is no adjustable parameter in the reaction part.
Therefore, 
nucleon-nucleus and $\alpha$-nucleus potentials are straightforwardly obtained 
from given structure inputs of diagonal and transition densities. 
The adopted channels of the MCC calculations are explained in Sec.~\ref{sec:results2}. 

The structure calculation of $\Be$ has been done 
by AMD with variation after parity and total angular momentum projections (VAP)
in Ref.~\cite{Kanada-Enyo:1999bsw}. The diagonal and transition densities obtained by AMD
have been used for the MCC calculation of the $\Be+p$ reaction in the 
previous work \cite{Kanada-Enyo:2019uvg}. We adopt the AMD results of $\Be$ 
as structure inputs of the present MCC calculations of the $\Be+p$ and $\Be+\alpha$ reactions.
For $\Mg$, we apply the AMD+VAP with fixed nucleon spins
in the same way Ref.~\cite{Kanada-Enyo:2020zpl}  for $^{28}$Si. 
Below, we briefly explain the AMD framework of the present calculation of $\Mg$. 
This calculation is 
an extension of the previous AMD calculation of $\Mg$ in Ref.~\cite{Kanada-Enyo:2011plo}.
For more details, the reader is referred to the previous works and references therein.

An AMD wave function of a mass-number $A$ nucleus is given by a Slater determinant of 
single-nucleon Gaussian wave functions as
\begin{eqnarray}
 \Phi_{\rm AMD}({\bvec{Z}}) &=& \frac{1}{\sqrt{A!}} {\cal{A}} \{
  \varphi_1,\varphi_2,\ldots,\varphi_A \},\label{eq:slater}\\
 \varphi_i&=& \phi_{{\bvec{X}}_i}\chi_i\tau_i,\\
 \phi_{{\bvec{X}}_i}({\bvec{r}}_j) & = &  \left(\frac{2\nu}{\pi}\right)^{3/4}
\exp\bigl[-\nu({\bvec{r}}_j-\bvec{X}_i)^2\bigr].
\label{eq:spatial}
\end{eqnarray}
Here ${\cal{A}}$ is the antisymmetrizer, and  $\varphi_i$ is
the $i$th single-particle wave function given by a product of
spatial ($\phi_{{\bvec{X}}_i}$), nucleon-spin ($\chi_i$), and isospin ($\tau_i$) 
wave functions.
In the present calculation of $\Mg$, we fix nucleon spin and isospin functions as 
spin-up and spin-down states of protons and neutrons. 
Gaussian centroid parameters $\{\bvec{X}_i\}$ for single-particle wave functions 
are treated as complex variational parameters independently for all nucleons.

In the model space of the AMD wave function, we perform energy variation after 
total-angular-momentum and parity projections (VAP). For each $J^\pi$ state, the variation is
performed with respect to the $J^\pi$-projected wave function $P^{J\pi}_{MK}\Phi_{\rm AMD}({\bvec{Z}})$
to obtain the optimum parameter set of Gaussian centroids $\{\bvec{X}_i\}$. 
Here $P^{J\pi}_{MK}$ is the total angular 
momentum and parity projection operator. In the energy variation, $K=0$ is taken for 
the $J^\pi=0^+$, $2^+$, and $4^+$ states in the $K^\pi=0^+_1$ ground-band, and $K=2$ is chosen for 
the $J^\pi=2^+$ and $3^+$ states in the $K^\pi=2^+$ side-band. After the energy variation of these states, 
we obtain five basis wave functions. To obtain final wave functions of $\Mg$, 
mixing of the five configurations (configuration mixing) and $K$-mixing 
are taken into account by diagonalizing the norm and Hamiltonian matrices.

In the present calculation of $\Mg$, 
the width parameter  $\nu=0.15$ fm$^{-2}$ is used.
The effective nuclear interactions of structure calculation for $\Mg$ are 
the MV1 (case 1) central force~\cite{TOHSAKI} supplemented by 
a spin-orbit term of the G3RS force~\cite{LS1,LS2}.
The Bartlett, Heisenberg, and Majorana parameters
of the MV1 force are $b=h=0$ and $m=0.62$, and the spin-orbit strengths are
$u_{I}=-u_{II}=2800$~MeV.
The Coulomb force is also included.
All these parameters of the Gaussian width and effective interactions 
are the same as those used 
in the previous studies of $\Mg$, $^{26}$Si, and $^{28}$Si of Refs.~\cite{Kanada-Enyo:2004ere,Kanada-Enyo:2011plo}.
A difference is the variational procedure.
The variation was done before the total angular momentum projection in the previous studies, but it
is done after the total angular momentum projection in the present AMD+VAP calculation.

\section{Energy levels, radii, and $B(E2)$ of target nuclei} \label{sec:results1}

\subsection{Structure of $\Mg$}

\begin{figure}[!h]
\includegraphics[width=8 cm]{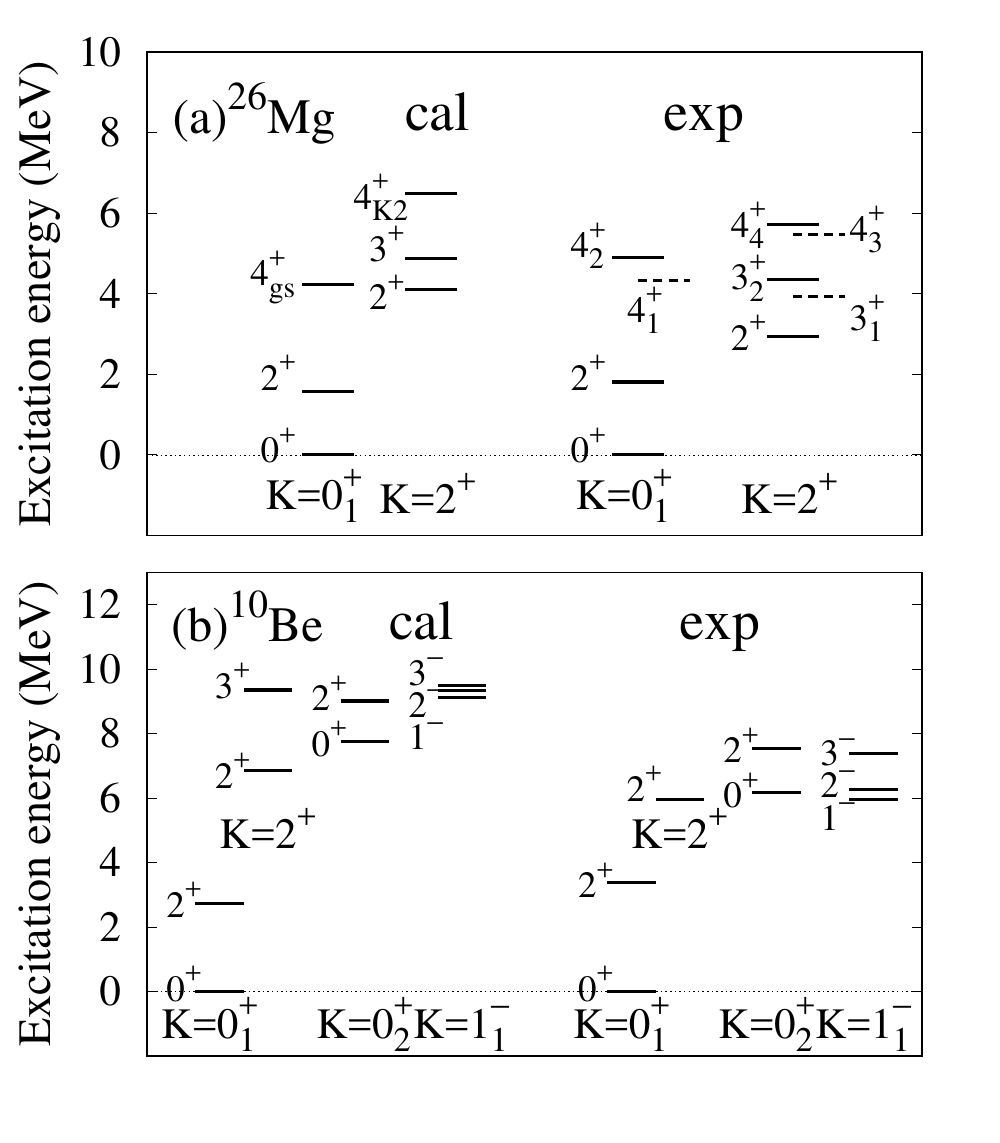}
  \caption{Energy levels of $\Mg$ and $\Be$.
(a) Calculated $\Mg$ levels of the $K^\pi=0^+_1$ ground-
and $K^\pi=2^+$ side-bands compared with the 
experimental levels. In the experimental spectra, 
candidate states for $3^+$ and $4^+$ states of band members 
are shown. (b) Calculated $\Be$ levels~\cite{Kanada-Enyo:1999bsw} of the $K^\pi=0^+_1$ ground-,  
$K^\pi=2^+$ side-, $K^\pi=0^+_2$ cluster-, and $K^\pi=1^-$ cluster-bands
are shown together with the observed energy levels. 
The experimental data are from Refs.~\cite{Basunia:2016egh,Tilley:2004zz}. 
  \label{fig:spe}}
\end{figure}

The ground and excited states of 
 $\Mg$ obtained after the diagonalization contain 
some amount of the configuration- and $K$-mixing, but they are approximately 
classified into the $K^\pi=0^+_1$ band built on the $0^+_1$ state 
and those in the $K^\pi=2^+$ band starting from the $2^+_2$ state.
In Fig.~\ref{fig:spe}(a),  
the calculated energy spectra are shown in comparison with the 
experimental spectra of candidate states for the  $K^\pi=0^+_1$ and $K^\pi=2^+$ band members.
The experimental $0^+_1$, $2^+_1(1.81)$, and $4^+_2(4.90)$ states are considered to belong the 
$K^\pi=0^+_1$ band, and the $2^+_2(2.94)$, $3^+_2(4.35)$, and $4^+_4(5.72)$ states are
tentatively assigned to the 
$K^\pi=2^+$ band from $\gamma$-decay properties~\cite{Nagel:1974}. However, there are 
other candidates such as the $3^+_1(3.94)$, $4^+_1(4.32)$, and $4^+_3(5.48)$ states 
in the same energy region.
We denote the theoretical states in the $K^\pi=0^+_1$ band as $J^\pi=\{0^+_1$, $2^+_1$, and $4^+_\textrm{gs}\}$
and those in the $K^\pi=2^+$ band as $J^\pi=\{2^+_{2}$, $3^+_{K2}$, and $4^+_{K2}\}$, 
and tentatively assign the $K^\pi=0^+_1$ band members to \{$0^+_1$, $2^+_1(1.81)$, $4^+_2(4.90)$\}
 and the $K^\pi=2^+$ band members to \{$2^+_{2}(2.94)$, $3^+_2(4.35)$, $4^+_4(5.72)$\}, 
though uncertainty remains in  assignments of $3^+$ and $4^+$ states. 

The root-mean-square~(rms) radii of proton~$(R_p$), neutron~$(R_n$), and matter~$(R_m$) distributions of 
the band-head  states of $\Mg$ are shown in Table~\ref{tab:radii-mg26}.
The $E\lambda$ transition strength $B(E\lambda)$ of the transition
$J^\pi_i\to J^\pi_f$ is given by the proton component of the matrix element $M_p$ as 
\begin{equation}
B(E\lambda;J^\pi_i\to J^\pi_f)\equiv \frac{1}{2J_i+1}|M_p|^2, 
\end{equation}
and its counter part (the neutron component $B_n(E\lambda)$) is given by $M_n$ 
as 
\begin{equation}
B_n(E\lambda;J^\pi_i\to J^\pi_f)\equiv \frac{1}{2J_i+1}|M_n|^2. 
\end{equation}
In Table~\ref{tab:BE2-mg26},  
the theoretical values of $B(E2)$ and $B_n(E2)$ obtained by AMD, and the 
observed $E2$ transition strengths are listed.

In each group of 
\{$0^+_1$, $2^+_1(1.81)$, $4^+_2(4.90)$\} and 
\{$2^+_2(2.94)$, $3^+_2(4.35)$,  $4^+_4(5.72)$\}, 
sequences of strong $\gamma$ transitions have been observed
and support the assignment of the $K^\pi=0^+_1$ and $K^\pi=2^+$
bands. 
However, possible state mixing between
the $4^+_1(4.32)$ and $4^+_2(4.90)$ states in the $K^\pi=0^+_1$ band is likely because of 
fragmentation of $E2$ transitions to the $2^+_1(1.81)$ state.
Moreover, an alternative assignment of 
the $K^\pi=2^+$ band composed of the $2^+_2(2.94)$, $3^+_1(3.94)$, and $4^+_3(5.48)$ states 
has been suggested~\cite{Durell:1972}. These experimental facts suggest that
collective natures of $3^+$ and $4^+$ states in these bands
may not be as striking as the rigid rotor picture. 

In the calculated result, the in-band transition strengths 
$B(E2;2^+_1\to 0^+_1)$ and $B(E2;4^+_\textrm{gs}\to 2^+_1)$ of the 
$K^\pi=0^+_1$ band
are remarkably large and in good agreement with the experimental data
for the  $0^+_1$, $2^+_1(1.81)$, and $4^+_2$(4.90) states.
For the $K^\pi=2^+$ band, the calculated $B(E2)$ values of the 
in-band transitions, $4^+_{K2}\to 2^+_2$, $4^+_{K2}\to 3^+_{K2}$, and $3^+_{K2}\to 2^+_2$, are 
a few times larger than the experimental $B(E2)$ of the $4^+_4\to 2^+_2$, $4^+_4\to 3^+_2$, and 
$3^+_2\to 2^+_2$ transitions, respectively, but relative ratios between three transitions are well 
reproduced by the calculation. 
It may indicate that 
the observed $2^+_2(2.94)$, $3^+_2(4.35)$, and $4^+_4(5.72)$ states possess
the $K^\pi=2^+$ band nature 
but the collectivity is somewhat quenched.  
It should be noted that the calculation shows significant inter-band transitions 
between the $K^\pi=0^+_1$ and $K^\pi=2^+$ bands such as
$4^+_\textrm{gs}\to 2^+_2$, which is consistent with the 
experimental $B(E2;4^+_2\to 2^+_2)$. 

Let us discuss the neutron component ($B_n(E2)$) of the transition strengths.
As seen in comparison of  $B_n(E2)$ and $B(E2)$, the neutron component 
is comparable to or even smaller than 
the proton component in most cases. 
Exceptions are 
the $2^+_2\to 0^+_1$ and $3^+_{K2}\to 2^+_1$ transitions, 
which show the neutron dominance indicating 
the predominant neutron excitation from the $K^\pi=0^+_1$ band to the 
$K^\pi=2^+$ band. It means the different isospin characters between 
two $2^+$ states, 
the $2^+_1$ state in the $K^\pi=0^+_1$ ground-band and the $2^+_2$ state
in the $K^\pi=2^+$ side-band. 
The former shows the approximately 
isoscalar feature and the latter has the neutron dominance character.

\begin{table}[!ht]
\caption{Calculated rms radii of proton ($R_p$), neutron ($R_n$), and matter ($R_m$) distributions 
of $\Mg$ and $\Be$~\cite{Kanada-Enyo:1999bsw}.
The experimental values of $R_p$ of the ground state are determined
from the experimental charge radii~\cite{Angeli2013}.
 \label{tab:radii-mg26}
}
\begin{center}
\begin{tabular}{llccccccccrrrrrrccccc}
\hline
\hline
 &    \multicolumn{3}{c}{AMD}&&exp\\	
 $^A Z(J^\pi)$ 	&	band	&	$R_p$ (fm)	&	$R_n$ (fm)	&	$R_m$ (fm)	&	$R_p$ (fm)	\\
$\Mg(0^+_1)$	&	$K^\pi=0^+_1$	&	3.10 	&	3.14 	&	3.12 	&	2.921(2)	\\
$\Mg(2^+_2)$	&	$K^\pi=2^+$	&	3.12 	&	3.15 	&	3.14 	&		\\
	&		&		&		&		&		\\
$\Be(0^+_1)$	&	$K^\pi=0^+_1$	&	2.50 	&	2.56 	&	2.54 	&	2.22(2)	\\
$\Be(2^+_2)$	&	$K^\pi=2^+$	&	2.60 	&	2.73 	&	2.68 	&		\\
$\Be(0^+_2)$	&	$K^\pi=0^+_2$	&	2.92 	&	3.17 	&	3.07 	&		\\
$\Be(1^-_1)$	&	$K^\pi=1^-$	&	2.75 	&	2.93 	&	2.86 	&		\\
\hline
\hline
\end{tabular}
\end{center}
\end{table}

\begin{table}[!ht]
\caption{The $E2$ transition strengths of $\Mg$. Theoretical values of 
proton ($B(E2)$) and neutron ($B_n(E2)$) components obtained by AMD, and 
the experimental $B(E2)$ values~\cite{Basunia:2016egh} are listed.
 \label{tab:BE2-mg26}
}
\begin{center}
\begin{tabular}{lrclrrrccccc}
\hline
\hline
{\ \ \ \ exp}&&\multicolumn{3}{l}{\ \ \ \ AMD} \\											
transition	&$	B(E2)				$&\ \ &	transition	&	$B(E2)$	&	$B_n(E2)$	\\
$2^+_1\to 0^+_1$	&$	61.3 	(	2.7 	)	$&&	$2^+_1\to 0^+_1$	&	63 	&	39 	\\
$2^+_2\to 0^+_1$	&$	1.8 	(	0.2 	)	$&&	$2^+_2\to 0^+_1$	&	0.8 	&	5.4 	\\
$4^+_1\to 2^+_1$	&$	21 	(	1 	)	$&&		&		&		\\
$4^+_2\to 2^+_1$	&$	64 	(	14 	)	$&&	$4^+_\textrm{gs}\to 2^+_1$	&	76 	&	58 	\\
$4^+_2\to 2^+_2$	&$	11 	(	3 	)	$&&	$4^+_\textrm{gs}\to 2^+_2$	&	8.8 	&	4.9 	\\
$4^+_2\to 3^+_2$	&$					$&&	$4^+_\textrm{gs}\to 3^+_{K2}$	&	39 	&	29 	\\
$4^+_3\to 2^+_1$	&$	5.0 	(	1.8 	)	$&&		&		&		\\
$4^+_3\to 3^+_1$	&$	55 	(	23 	)	$&&		&		&		\\
$4^+_4\to 2^+_1$	&$					$&&	$4^+_{K2}\to 2^+_1$	&	11.4 	&	3.5 	\\
$4^+_4\to 2^+_2$	&$	7.8 	(	2.3 	)	$&&	$4^+_{K2}\to 2^+_2$	&	22 	&	11 	\\
$4^+_4\to 3^+_1$	&$	1.8 	(	0.9 	)	$&&		&		&		\\
$4^+_4\to 3^+_2$	&$	14 	(	6 	)	$&&	$4^+_{K2}\to 3^+_{K2}$	&	39 	&	22 	\\
$3^+_2\to 2^+_1$	&$	0.3 	(	0.2 	)	$&&	$3^+_{K2}\to 2^+_1$	&	1.5 	&	9.4 	\\
$3^+_2\to 2^+_2$	&$	41 	(	18 	)	$&&	$3^+_{K2}\to 2^+_2$	&	114 	&	66 	\\
\hline
\end{tabular}
\end{center}
\end{table}

\subsection{Structure of $\Be$}

In the AMD calculation of $\Be$, the $2\alpha+nn$ cluster structures are obtained in the ground and 
excited states as discussed in Ref.~\cite{Kanada-Enyo:1999bsw}.
The $K^\pi=0^+_1$ ground- and $K^\pi=2^+$ side-bands are constructed. 
In addition, the $K^\pi=0^+_2$ and $K^\pi=1^-_1$ cluster-bands are obtained.
The energy spectra of $\Be$ are shown in Fig.~\ref{fig:spe}(b).
The calculated energy levels are in reasonable agreement with the experimental spectra.
The calculated rms proton, neutron, and matter radii of the band-head states
are given in Table~\ref{tab:radii-mg26}. The $0^+_2$($K^\pi=0^+_2$)  and $1^-_1$($K^\pi=1^-$)  states of the 
cluster-bands have relatively larger radii
compared to the $0^+_1$($K^\pi=0^+_1$) and $2^+$($K^\pi=2^+$) states
because of the developed cluster structure.

The calculated result of the transition strengths and matrix elements of 
the monopole (IS0), dipole (IS1), $E2$, and $E3$ transitions are summarized in 
Table~\ref{tab:BE2-be10}. The $M_n/M_p$ ratio and the 
isoscalar component $B_{p+n}\equiv|M_p+M_n|^2/(2J_i+1)$ of the transition strength 
are also given in the table.
For experimental data, the $E2$ transition strengths and matrix elements observed for  
$\Be$ and those for the mirror nucleus $^{10}\textrm{C}$ are listed. 
The experimental $M_n$ value of $\Be$ is
evaluated from the mirror transition assuming the mirror symmetry (no charge effect for $A=10$ nuclei).
One of the striking features is that, in many transitions of $\Be$,  
the neutron component is dominant compared to the proton component
because of contributions of valence neutrons around the $2\alpha$ cluster.
An exception is the $2^+_1\to 0^+_1$ transition in the $K^\pi=0^+_1$ ground-band having  
the isoscalar nature of nearly equal proton and neutron components, which are generated by
the $2\alpha$ core rotation. 

As a result, isospin characters of  the ground-band $2^+_1$ state and the side-band $2^+_2$ state
are quite different from each other. 
The former has the isoscalar feature and the latter shows the neutron dominance character.
This is similar to the case of $\Mg$ and can be a general feature
of $N=Z+2$ system having a $N=Z$ core with prolate deformation.
The ground-band $2^+$ state is constructed by the $K=0$ rotation of 
the core part with the isoscalar prolate deformation, 
whereas the side-band $2^+$ state is described by the $K=2$ rotation of valence neutrons 
around the prolate core.

\begin{table*}[ht]
\caption{Transition strengths and matrix elements 
of the isoscalar monopole (IS0) and dipole (IS1), and $E\lambda$ transitions. 
The calculated values are the isoscalar ($p+n$), proton, and neutron components of the transition strengths, 
the proton and neutron transition matrix elements, and the $M_n/M_p$ ratio 
obtained by AMD~\cite{Kanada-Enyo:1999bsw}. The experimental values are $E2$ transition strengths of 
$\Be$ and $^{10}\textrm{C}$, and $M_p$,  $M_n$, and $M_n/M_p$
from Ref.~\cite{Tilley:2004zz}. 
$^{b}$ The empirical values of $M_n$ and $M_n/M_p$ evaluated 
from the mirror transition assuming the mirror symmetry.
 \label{tab:BE2-be10}
}
\begin{center}
\begin{tabular}{ccccccccc}
\hline
\hline
&\multicolumn{5}{c}{AMD} \\											
	&	$B_{p+n}(\textrm{IS0})$	&	$B_p(\textrm{IS0})$	&	$B_n(\textrm{IS0})$	&	$M_p$	&	$M_n$	&	$M_n/M_p$	\\
$0^+_2 \to 0^+_1$	&	12.7 	&	1.5 	&	5.4 	&	1.2 	&	2.3 	&	1.89 	\\
	&	$B_{p+n}(E2)$	&	$B(E2)$	&	$B_{n}(E2)$	&		&		&		\\
$2^+_1 \to 0^+_1$	&	41 	&	11.6 	&	8.9 	&	7.6 	&	6.7 	&	0.88 	\\
$2^+_2 \to 0^+_1$	&	1.7 	&	0.2 	&	3.2 	&$	-1.0 	$&	4.0 	&$	-3.9 	$\\
$2^+_3 \to 0^+_1$	&	1.5 	&	0.1 	&	0.7 	&	0.8 	&	1.9 	&	2.5 	\\
$2^+_3 \to 0^+_2$	&	280 	&	34 	&	118 	&	13.1 	&	24.3 	&	1.85 	\\
$0^+_2 \to 2^+_1$	&	6.0 	&	0.6 	&	2.9 	&	0.7 	&	1.7 	&	2.3 	\\
	&	$B_{p+n}(\textrm{IS1})$	&	$B_p(\textrm{IS1})$	&	$B_n(\textrm{IS1})$	&		&		&		\\
$1^-_1 \to 0^+_1$	&	6.0 	&	1.0 	&	2.1 	&	1.7 	&	2.5 	&	1.46 	\\
	&	$B_{p+n}(E3)$	&	$B(E3)$	&	$B_{n}(E3)$	&		&		&		\\
$3^-_1 \to 0^+_1$	&	70 	&	1.3 	&	53 	&	3.0 	&	19.2 	&	6.4 	\\
\hline
&\multicolumn{5}{c}{exp} \\											&		
	&		$B(E2)$	&	$B(^{10}\textrm{C};E2)$	&	$M_p$	&	$M_n$	&	$M_n/M_p$	\\
$2^+_1 \to 0^+_1$	&		&	10.2(1.0)	&	12.2(1.9)	&	7.2(0.4)	&	7.8(0.6)$^{b}$	&	1.08$^{b}$ 	\\
$0^+_2 \to 2^+_1$	&		&	3.2(1.2)	&		&	1.8(0.4)	&		&		\\
\hline
\end{tabular}
\end{center}
\end{table*}

\section{$p$ and $\alpha$ scattering}  \label{sec:results2}

In order to reduce model ambiguity of structure inputs, 
we perform fine tuning of the theoretical transition densities $\rho^\textrm{tr}(r)$ 
by multiplying overall factors as $\rho^\textrm{tr}(r)\to f^\textrm{tr}\rho^\textrm{tr}(r)$ 
to fit the observed $B(E\lambda)$ data, 
and utilize the renormalized transition densities $f^\textrm{tr}\rho^\textrm{tr}(r)$ 
for the MCC calculations. 
For each system of $\Mg$ and $\Be$, we first describe the scaling factors $f^\textrm{tr}$ and show the renormalized
transition densities and form factors. Then, 
we investigate $p$ and $\alpha$ scattering cross sections 
with the MCC calculations using the renormalized AMD densities
to clarify to transition properties of excited states, in particular, their 
isospin characters.

\subsection{Transition properties of $\Mg$}

The transition matrix elements ($M_p$ and $M_n$) and
the scaling factors ($f^\textrm{tr}_p$ and $f^\textrm{tr}_p$) for the renormalization of 
transition densities are listed in Table~\ref{tab:MnMp}. 
Theoretical values before and after the renormalization are shown together with the 
experimental $M_p$ and $M_n$ values used for fitting.

For renormalization of the
$2^+_1\to 0^+_1$ and $2^+_2\to 0^+_1$ transitions,  
we determine the scaling factor $f^\textrm{tr}_p$ of the proton
transition density to fit the experimental $M_p$ values measured by $\gamma$ decays,
and $f^\textrm{tr}_n$ of the neutron transition density by fitting 
the experimental $M_n$ values, which are 
evaluated from the mirror transitions with a correction factor 0.909 of charge 
effects~\cite{Brown:1977vxz} in the same way as Ref.~\cite{Alexander:1982sph}.
In order to see the sensitivity of the cross sections to the isospin character 
of the $K^\pi=2^+$ side-band, 
we also consider two optional sets (case-1 and case-2) 
of ($f^\textrm{tr}_p$,$f^\textrm{tr}_n$) for the $2^+_2\to 0^+_1$ transition, which are 
discussed in details later in Sec.~\ref{sec:mg26-reaction}.  

For $4^+\to 0^+_1$ transitions, $B(E4)$ has not been measured $\gamma$ rays
but the transition strengths have been evaluated by inelastic scattering experiments.
In Table~\ref{tab:BE4-mg26}, we list the transition strengths (or rates) of the $2^+\to 0^+_1$ and 
$4^+\to 0^+_1$ transitions: electric transition strengths 
$B(E\lambda)$ obtained with $(e,e')$ data~\cite{Lees1974},  
$\alpha$ inelastic transition rates $B_{\alpha,\alpha'}$ evaluated from the
$(\alpha,\alpha')$ study~\cite{VanDerBorg:1981qiu}, 
and $p$ inelastic transition rates $B_{p,p'}$ from the 
$(p,p')$ reaction~\cite{Alons:1981rrm,Zwieglinski:1983zz}.
Note that hadron scattering probes not only the proton but also the neutron components of transitions rates.
In the present calculation, we adopt the $B(E4)$ values of the $4^+_2(4.90)$ and $4^+_4(5.72)$ states
obtained from the $(e,e')$ experiments to
determine $f^\textrm{tr}_p$ for
the theoretical $4^+_\textrm{gs}$ and $4^+_{K2}$ states, respectively. 
For $f^\textrm{tr}_n$ of the neutron transition density, 
we use the same values as $f^\textrm{tr}_p$. 

\begin{figure}[!h]
\includegraphics[width=7 cm]{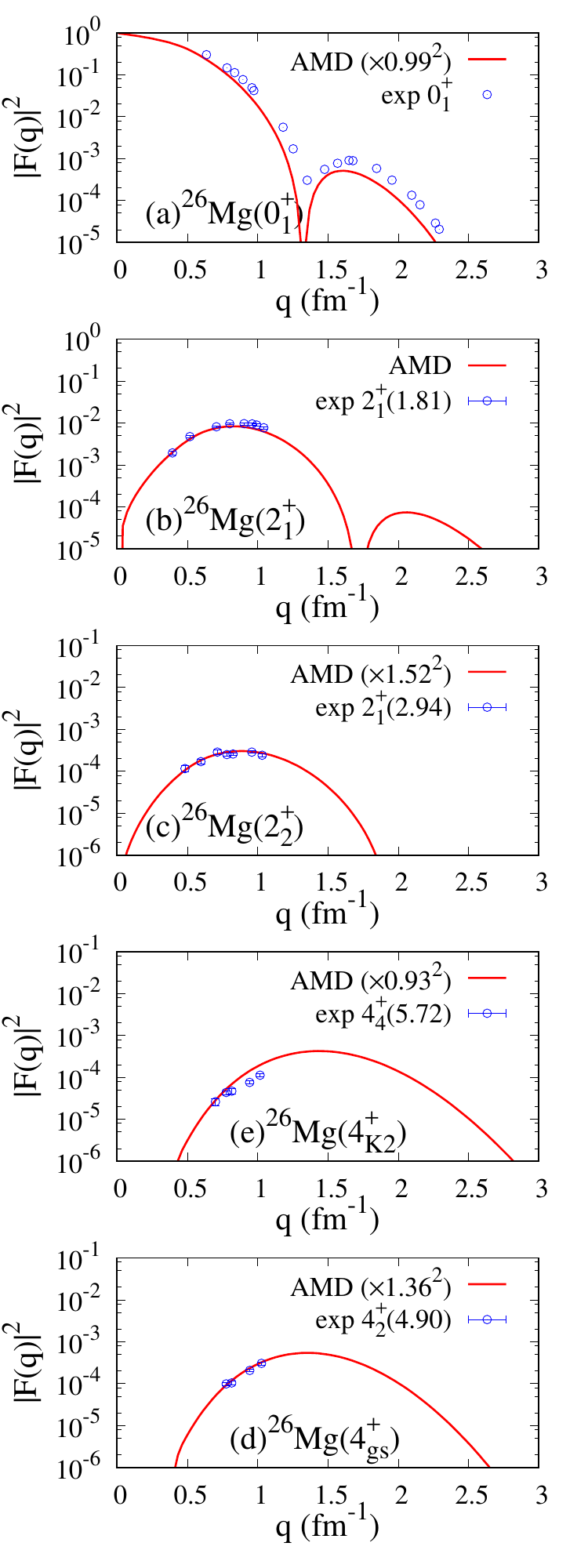}
  \caption{Charge form factors of $\Mg$. 
The inelastic form factors $F(q)$ obtained by AMD are renormalized by
$f^\textrm{tr}_p$ given in Table~\ref{tab:MnMp}. 
The results of the $0^+_1$, $2^+_1$, $2^+_2$, $4^+_\textrm{gs}$, and $4^+_{K2}$ 
states are compared with the experimental data~\cite{Lees1974} of the $0^+_1$, $2^+_1$(1.81~MeV), $2^+_2$(2.94~MeV),  $4^+_2$(4.90~MeV), and $4^+_4$(5.72~MeV) states, respectively.
  \label{fig:form-mg26}}
\end{figure}

\begin{figure}[!h]
\includegraphics[width=8 cm]{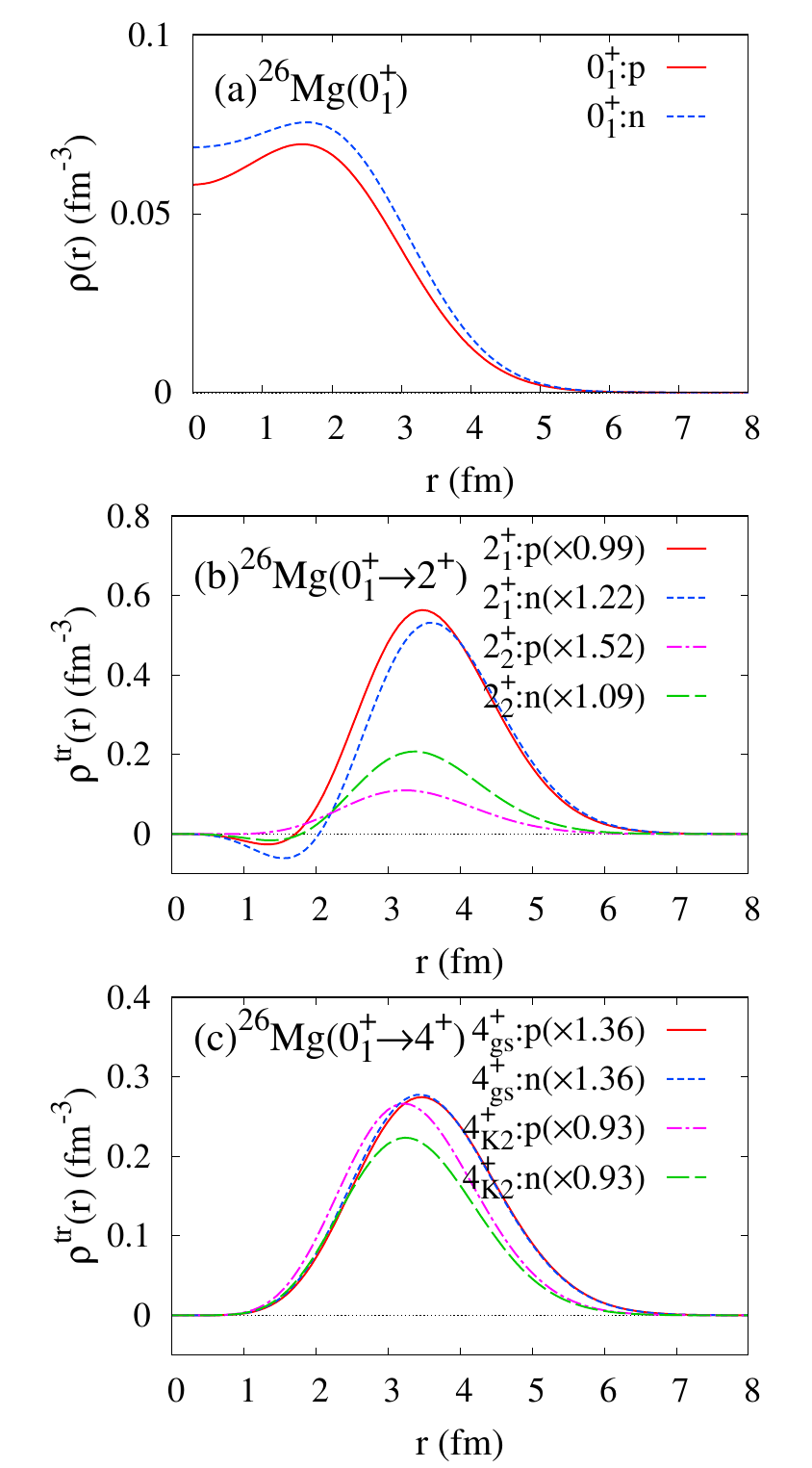}
  \caption{Proton and neutron diagonal and transition densities of  $\Mg$.
(a) The diagonal densities of the $0^+_1$ state.
(b) The renormalized transition densities from the $0^+_1$ state to the $2^+_1$ and $2^+_2$ states. 
(c) The renormalized transition densities from the $0^+_1$ state to the $4^+_\textrm{gs}$ 
and $4^+_{K2}$ states. 
  \label{fig:trans-mg26}}
\end{figure}

Figure~\ref{fig:form-mg26} shows 
the calculated elastic and inelastic form factors of $\Mg$ in comparison with
the experimental data.
The data are well reproduced by the renormalized form factors of AMD.
In Fig.~\ref{fig:trans-mg26}, we show 
the diagonal densities and the renormalized transition densities.
In the ground-band transitions, $0^+_1\to 2^+_1$ and $0^+_1\to 4^+_\textrm{gs}$, 
the proton and neutron transition densities are almost the same as each other showing 
the isoscalar nature of those excitations in the $K^\pi=0^+_1$ ground-band. 
In the $0^+_1\to 2^+_2$ excitation 
to the $K^\pi=2^+$ side-band, the neutron transition density is about twice larger than
the proton one showing the neutron dominance, while 
the transition densities of $0^+_1\to 4^+_\textrm{K2}$ show the isoscalar nature. 
In radial behavior of the transition densities to the $2^+_1$  and $2^+_2$ states, 
one can see that the peak position slightly shifts to the inner region 
in the $0^+_1\to 2^+_2$ transition compared to the
$0^+_1\to 2^+_1$ transition.

\begin{table}[ht]
\caption{
The transition matrix elements ($M_p$ and $M_n$ in the unit of  fm$^\lambda$)
 and the ratios ($M_n/M_p$), and 
the scaling factors ($f^\textrm{tr}_p$ and $f^\textrm{tr}_n$) for the renormalization of 
transition densities.
For use of the default MCC calculations, 
the scaling factors $f^\textrm{tr}_p$ and  $f^\textrm{tr}_n$ 
for the proton and neutron components are determined 
so as to fit the experimental $M_p$ and $M_n$ values, respectively. 
Theoretical values before (theor.) and after (default MCC) the renormalization are shown together with the 
experimental values~\cite{Basunia:2016egh,Tilley:2004zz}. 
For the $E2;2^+_2\to 0^+_1$ transition of $\Mg$, 
two optional sets (case-1 and case-2) of $f^\textrm{tr}_{p,n}$
are considered in addition to the default scaling. 
$^{(a)}$The $M_n$ values of $\Mg$ are estimated from the mirror transitions with correction  
0.909 of charge effects. 
This correction was given for $A=18$ in Ref.~\cite{Brown:1977vxz}
and used arbitrarily for $A=26$ nuclei as done in Ref.~\cite{Alexander:1982sph}.
$^{(b)}$The $M_n$ value of $\Be$ from the mirror transition assuming 
the mirror symmetry  (no charge effect) for $A=10$ nuclei. 
 \label{tab:MnMp}
}
\begin{center}
\begin{tabular}{cccccccccc}
\hline
\hline										
	&	$M_p$	&	$M_n$	&	$M_n/M_p$	&	$f^\textrm{tr}_p$	&	$f^\textrm{tr}_n$	\\
\multicolumn{2}{c}{$\Mg(E2:2^+_1\to 0^+_1)$}											\\
exp	&	17.5(0.4)	&	17.0(1.0)$^{(a)}$&	0.97 	&		&		\\
theor.	&	17.7 	&	13.9 	&	0.79 	&	1	&	1	\\
MCC(default)	&	17.5 	&	17.0 	&	0.97$^{(a)}$ 	&	0.99 	&	1.22 	\\
&\\ \multicolumn{2}{c}{$\Mg(E2:2^+_2\to 0^+_1)$}											\\
exp 	&	3.0(0.1)	&	5.7(0.6)$^{(a)}$&	1.90 $^{(a)}$&		&		\\
theor.	&	2.0 	&	5.2 	&	2.65 	&	1	&	1	\\
MCC(default)	&	3.0 	&	5.7 	&	1.90 	&	1.52 	&	1.09 	\\
MCC(case-1)	&	3.0 	&	7.9 	&	2.65 	&	1.52 	&	1.52 	\\
MCC(case-2)	&	4.3 	&	4.3 	&	1.00 	&	2.20 	&	0.83 	\\
&\\ \multicolumn{2}{c}{$\Mg(E4:4^+_2\to 0^+_1)$}											\\
exp $(e,e')$	&	161(21)	&		&		&		&		\\
theor.	&	119 	&	118 	&	0.99 	&	1	&	1	\\
MCC(default)	&	162 	&	161 	&	0.99 	&	1.36	&	1.36	\\
&\\ \multicolumn{2}{c}{$\Mg(E4:4^+_4\to 0^+_1)$}											\\
exp $(e,e')$	&	114(20)	&		&		&		&		\\
theor.	&	123 	&	105 	&	0.85 	&	1	&	1	\\
MCC(default)	&	114 	&	97 	&	0.85 	&	0.93	&	0.93	\\
&\\ \multicolumn{2}{c}{$\Be(E2:2^+_1\to 0^+_1)$}											\\
exp	&	7.2(0.4)	&	7.8(0.6)$^{(b)}$	&	1.08 $^{(b)}$	&		&		\\
theor.	&	7.6 	&	6.7 	&	0.88 	&	1	&	1	\\
MCC(default)	&	7.2 	&	7.8 	&	1.09 	&	0.94 	&	1.17 	\\
\hline
\end{tabular}
\end{center}
\end{table}
\begin{table}
\caption{The $E2;2^+\to 0^+_1$ and $E4;4^+\to 0^+_1$ transition strengths of 
$\Mg$ evaluated from the $(e,e')$, $(p,p')$, and $(\alpha,\alpha')$ reactions. 
Electric transition strengths $B_{e,e'}(E\lambda)$ obtained by the $(e,e')$ experiments~\cite{Lees1974},
$\alpha$ inelastic transition rates $B_{\alpha,\alpha'}$ evaluated 
by $(\alpha,\alpha')$ at $E_\alpha=120$~MeV~\cite{VanDerBorg:1981qiu}, 
and $p$ inelastic transition rates $B_{p,p'}$ by 
$(p,p')$ at  $E_p=40$~MeV~\cite{Zwieglinski:1983zz} and $E_p=24$~MeV~\cite{Alons:1981rrm} are shown 
together with the theoretical values of the 
proton and neutron components, $B(E\lambda)$ and $B_n(E\lambda)$, of the strengths.
The units are
$\textrm{fm}^{2\lambda}$.
 \label{tab:BE4-mg26}
}
\begin{center}
\begin{tabular}{lcccccccc}
\hline
\hline
&\multicolumn{4}{c}{exp}&\multicolumn{2}{c}{AMD} \\						
$J^\pi[E_x]$	&	$B_{e,e'}(E\lambda)$	&	$B_{\alpha,\alpha'}$	&	$B_{p,p'}$	&	$B_{p,p'}$		&	$B(E\lambda)$	&	$B_n(E\lambda)$	\\
	&	Ref.~\cite{Lees1974}	&	Ref.~\cite{VanDerBorg:1981qiu}	&	Ref.~\cite{Zwieglinski:1983zz} 	&	
	Ref.~\cite{Alons:1981rrm}		\\
$2^+_1(1.81)$	&	53.2(3.2)	&	55	&	46(1)	&	37(2)	&	63 	&	39 	\\
$2^+_2(2.94)$	&	1.3(0.3)	&	7.8	&	6.6(0.2)	&	5.6(0.6)	&	0.8 	&	5.4 	\\
$4^+_1(4.32)$	&		&	9.7	&	11.0(0.8)	&	4.5(0.5)	&		&		\\
$4^+_2(4.90)$	&	29(8)	&	11.5	&	21(1)	&	10.6(0.9)	&	15.7	&	15.5	\\
$4^+_3(5.48)$	&	3.8	&	11.5	&	7.7(0.6)	&	4.4(0.6)	&		&		\\
$4^+_4(5.72)$	&	14(6)	&	5.2	&	4(0.2)	&		&	16.8	&	12.2	\\
\hline
\end{tabular}
\end{center}
\end{table}

\subsection{$\Mg+p$ and $\Mg+\alpha$ reactions}\label{sec:mg26-reaction}
Using the AMD densities of  $\Mg$, 
we perform the MCC calculations of $p$ scattering at $E_p=24$, 40, 60, and 100~MeV and $\alpha$
scattering at $E_\alpha=104$, 120, and 400~MeV. 
For coupled channels, we take into account 
the $0^+_1$, $2^+_1$, $2^+_2$, $4^+_\textrm{gs}$, and $4^+_{K2}$
states and $\lambda=2$ and 4 transitions between them. To see coupled channel~(CC) effects, 
we also calculate one-step cross sections 
with distorted wave Born approximation~(DWBA).
The experimental excitation energies of the $0^+_1$, $2^+_1$, $2^+_2$, $4^+_2$, and $4^+_{4}$ states 
are used in the 
reaction calculations. 
For the transitions of  $2^+_{1} \to 0^+_1$, $4^+_\textrm{gs} \to 0^+_1$, 
$2^+_{2} \to 0^+_1$, and $4^+_{K2} \to 0^+_1$, 
the renormalized transition densities are used as explained previously. 
For other transitions, 
we use theoretical transition densities without  renormalization.

\begin{figure*}[!h]
\includegraphics[width=18 cm]{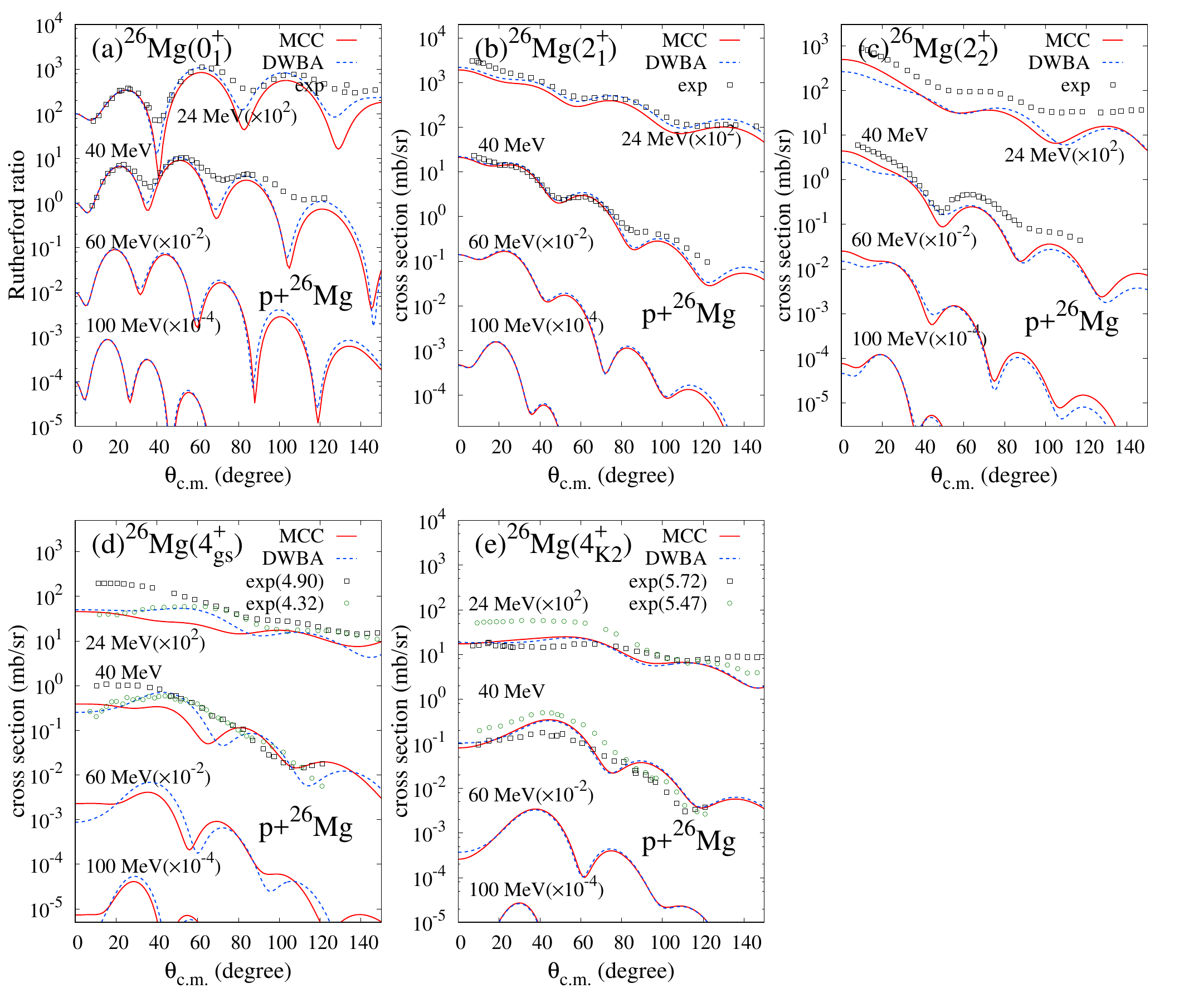}
  \caption{
Cross sections of $p$ elastic and inelastic scattering off $\Mg$ at 
$E_p=24$, 40, 60, and 100~MeV. The results obtained by the MCC and DWBA calculations 
are shown by red solid and blue dashed lines, respectively. 
Experimental data are cross sections at 
$E_p=24$~MeV~\cite{Alons:1981rrm} and 
$40$~MeV~\cite{Zwieglinski:1983zz}.
The panels (a), (b), and (c) show the calculated and experimental cross sections of the $0^+_1$, $2^+_1$, and $2^+_2$ states, 
respectively. The panel (d) shows the calculated $4^+_\textrm{gs}$ cross sections together with 
the data observed for the $4^+_2$(4.90~MeV) and $4^+_1$(4.32~MeV) states.
 The panel (e) shows the calculated $4^+_{K2}$ cross sections compared with 
the data observed for the $4^+_4$(5.47~MeV) and $4^+_3$(5.72~MeV) states.
  \label{fig:cross-mg26p}}
\end{figure*}

\begin{figure*}[!h]
\includegraphics[width=18. cm]{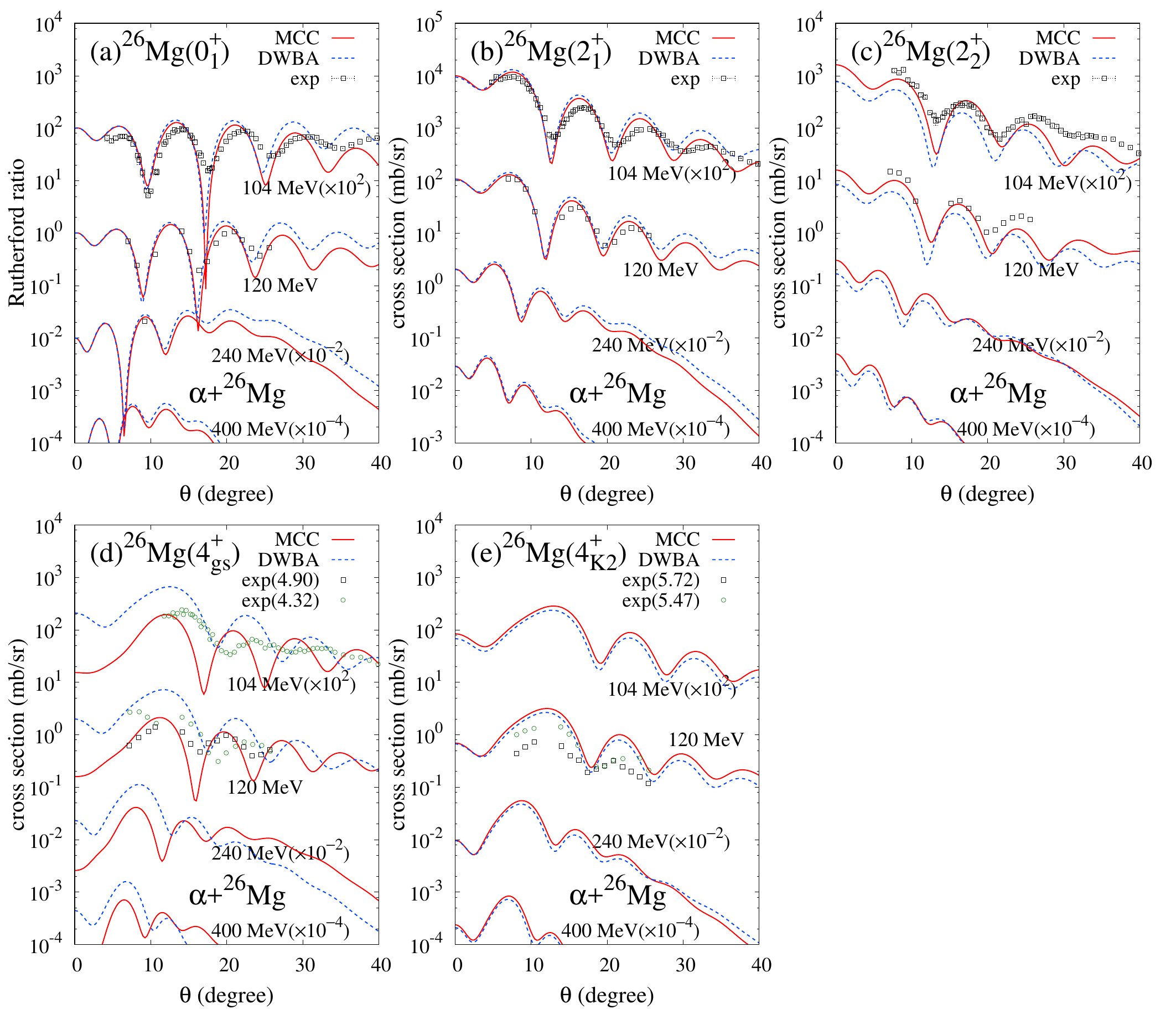}
  \caption{
Same as Fig.~\ref{fig:cross-mg26p} but for $\alpha$ scattering at $E_\alpha=104$, 120, 240, and 400~MeV. 
Experimental data at 
$E_\alpha=104$~MeV~\cite{Rebel:1972nip} and 
$E_\alpha=120$~MeV~\cite{VanDerBorg:1981qiu} are shown.
  \label{fig:cross-mg26a}}
\end{figure*}

\begin{figure*}[!h]
\includegraphics[width=12. cm]{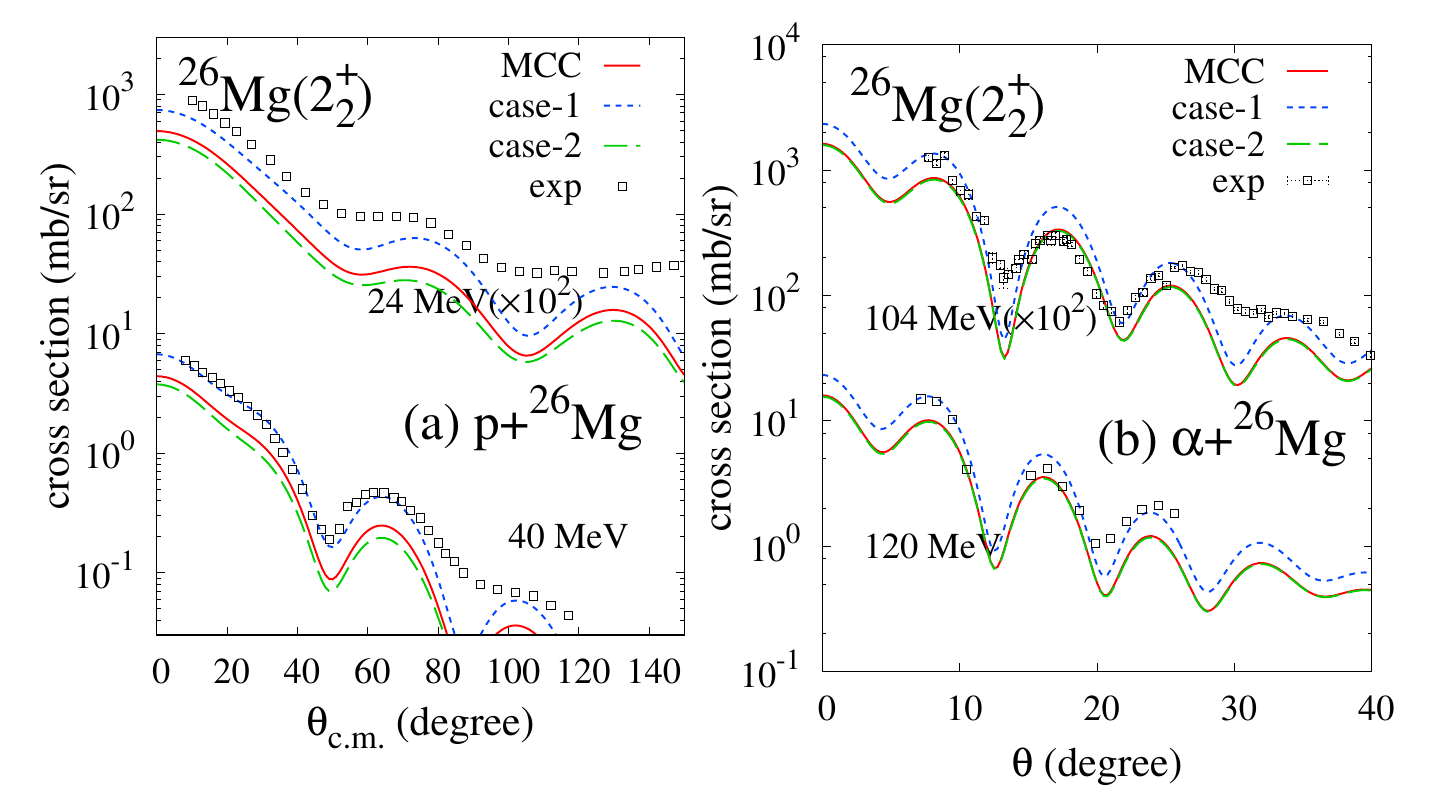}
  \caption{
$2^+_2$ cross sections of the $\Mg+p$ and $\Mg+\alpha$ reactions calculated by MCC
using the renormalized transition densities with the default, case-1, and case-2 scaling. 
(a) ($p,p')$ cross sections at $E_p=24$ and $40$~MeV and 
(b) $(\alpha,\alpha')$ cross sections at $E_\alpha=104$ and $120$~MeV.
The experimental data of $(p,p')$ are from 
Refs.~\cite{Alons:1981rrm,Zwieglinski:1983zz}, those of $(\alpha,\alpha')$ are 
from Refs.~\cite{Rebel:1972nip,VanDerBorg:1981qiu}.
  \label{fig:cross-mg26p-m}}
\end{figure*}

The  MCC and DWBA results of the $\Mg+p$ reaction are shown in Fig.~\ref{fig:cross-mg26p}
together with the experimental cross sections at $E_p=24$ and  40~MeV, 
and those of the $\Mg+\alpha$ reaction are shown in Fig.~\ref{fig:cross-mg26a} with
the experimental cross sections at $E_\alpha=104$ and 120~MeV.
As shown in Fig.~\ref{fig:cross-mg26p}(a), 
the MCC calculation reproduces well the $p$ elastic cross sections at $E_p=24$ and 40~MeV.
It should be commented that spin-orbit interaction, which is omitted in the present reaction 
calculation,
may smear the deep dip structure of the calculated cross sections. 
The calculation also describes the experimental data of $\alpha$ elastic scattering at 
$E_\alpha=104$ and 120~MeV (Fig.~\ref{fig:cross-mg26a}(a)).
 
For the ground-band $2^+_1$ state, the MCC calculation successfully
reproduces the amplitudes and also the diffraction patterns of the 
$(p,p')$ and $(\alpha,\alpha')$ cross sections.
For the inelastic scattering to the side-band 
$2^+_2$ state, the calculation reasonably describes the $(\alpha,\alpha')$ data but somewhat underestimates the 
$(p,p')$ data. 
Comparing the DWBA and MCC results, one can see that CC
effects are minor in the $2^+_1$ and $2^+_2$ cross sections of $p$ scattering
and $2^+_1$ cross sections of $\alpha$ scattering
but give a significant contribution to 
the $2^+_2$ cross sections of low-energy $\alpha$ scattering. 

For the $4^+_\textrm{gs}$ and $4^+_{K2}$ states, agreements with the experimental 
$(p,p')$ cross sections 
are not enough satisfactory to discuss whether the present assignment of $4^+$ states 
is reasonable (Fig.~\ref{fig:cross-mg26p}(d) and Fig.~\ref{fig:cross-mg26p}(e)).
For the $(\alpha,\alpha')$ processes, the experimental cross sections 
observed for the $4^+_2$(4.90~MeV) state are reproduced well by the MCC result, which shows
large suppression by the CC effect (Fig.~\ref{fig:cross-mg26a}(d)). 
For the $4^+_{K2}$ cross sections, the MCC calculation obtains almost no suppression 
by the CC effect
and significantly overestimates the $(\alpha,\alpha')$ data for the 
$4^+_4$(5.72~MeV) states.
We can state that $p$ and $\alpha$ inelastic processes to low-lying $4^+$ states 
are not as simple as a theoretical description with the $4^+_\textrm{gs}$ and $4^+_{K2}$ states. 
Instead, they may be affected by 
significant state mixing and channel coupling, which are beyond the 
present AMD calculation. This indication is consistent with the $\gamma$ decay properties. 

Let us discuss isospin properties of the $2^+_1$ and $2^+_2$ states with further detailed analysis of the
inelastic cross sections. As shown previously, 
the MCC calculation gives good reproduction 
of the $2^+_1$ cross sections in describing 
the peak and dip structures of the $(p,p')$ data at $E_p=40$~MeV and $(\alpha,\alpha')$
data at $E_\alpha$=120~MeV (Fig.~\ref{fig:cross-mg26p}(b) and Fig.~\ref{fig:cross-mg26a}(b)). 
For the $2^+_2$ state, 
it describes the diffraction patterns of the $(p,p')$ data but somewhat 
underestimates absolute amplitudes of the cross sections. 
In order to discuss possible uncertainty in the neutron strength (or the $M_n/M_p$ ratio) 
of the 
$0^+_1\to 2^+_2$ transition, 
we consider here optional choices of the renormalization
of the transition densities by changing the scaling factors
 ($f^\textrm{tr}_p$,$f^\textrm{tr}_n$) for this transition
from the default values 
$(f^\textrm{tr}_p,f^\textrm{tr}_n)=(1.52,1.09)$. 
The values of $(f^\textrm{tr}_p,f^\textrm{tr}_n)$, $M_p$, $M_n$, and $M_n/M_p$ 
for these two choices are listed in Table~\ref{tab:MnMp}.  
In the case-1, we choose the same scaling for the proton and neutron parts as 
$(f^\textrm{tr}_p,f^\textrm{tr}_n)=(1.52,1.52)$.
In this case, the neutron transition density 
is enhanced by 40\% from the default MCC calculation
(the neutron transition strengths is enhanced by a factor of two).   
The case-2 choice is  $(f^\textrm{tr}_p, f^\textrm{tr}_n)=(2.20, 0.83)$, which 
corresponds to an  
assumption of the isoscalar transition $M_p=M_n$ keeping the isoscalar component 
$M_p+M_n$ unchanged. In these two optional cases, other transitions are the same as the default calculation. 
In Fig.~\ref{fig:cross-mg26p-m}, 
we show the  $2^+_2$ cross sections obtained by MCC with the case-1 and case-2 choices. 
In the case-1 calculation, one can see that the 40\% increase of the 
neutron transition density significantly enhances the $(p,p')$ cross sections and  
slightly raises the $(\alpha,\alpha'$) cross sections. As a result, the calculation 
well reproduces the $(p,p')$ cross sections, in particular, at $E_p=40$~MeV
and also obtains a better result for the $(\alpha,\alpha')$ cross sections.
In the case-2 calculation (isoscalar assumption), the result for $(p,p')$ cross sections 
becomes somewhat worse, and that for $(\alpha,\alpha')$ cross sections is unchanged.
This result indicates that the $(p,p')$ process sensitively probes the dominant neutron component 
of the $0^+_1\to 2^+_2$ transition and 
the $(\alpha,\alpha')$ process can probe the isoscalar component as expected. 
In the present analysis, the case-1 calculation is favored to describe the $2^+_2$ cross sections 
in both the $(p,p')$ and $(\alpha,\alpha')$ processes. 
This analysis supports the case-1 prediction 
for the $0^+_1\to 2^+_2$ transition of 
the neutron transition matrix $M_n\sim 8$~fm$^2$ 
(the squared 
ratio $|M_n/M_p|^2\sim 7$).

\subsection{Transition properties of $\Be$}

For $\Be$, experimental information of $B(E\lambda)$ is limited. 
For the transition from the $0^+_1$ state,  
the available data
are the observed values of $B(E2;2^+_1\to 0^+_1)$ and its mirror transition, 
with which we adjust the scaling factors of the renormalization.
The transition matrix elements ($M_p$ and $M_n$) and
the scaling factors ($f^\textrm{tr}_p$ and $f^\textrm{tr}_p$) of $2^+_1\to 0^+_1$ in $\Be$
are given in Table~\ref{tab:MnMp}. 
Theoretical values before and after the renormalization are shown together with the 
experimental values used for fitting.
For other transitions, theoretical transition densities without the renormalization are used 
for the MCC calculation.

\begin{figure}[!h]
\includegraphics[width=6 cm]{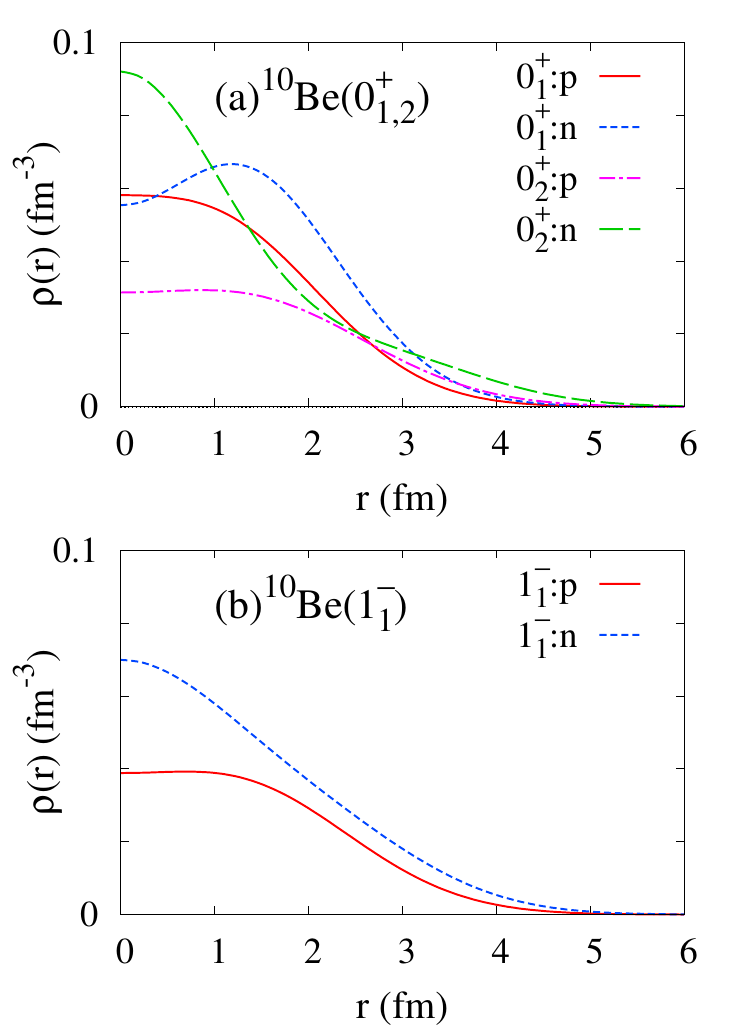}
  \caption{
Diagonal densities of $\Be$.
The proton and neutron densities of the (a) $0^+_{1,2}$ and (b) $1^-_1$ states.
  \label{fig:dens-be10}}
\end{figure}

Figure~\ref{fig:dens-be10} shows calculated diagonal densities of $\Be$.
Compared to the ground state, 
the $0^+_2$($K^\pi=0^+_2$) and $1^-_1$($K^\pi=1^-$) states 
show longer tails of the proton and neutron diagonal densities
because of the developed cluster structures. 

\begin{figure}[!h]
\includegraphics[width=6 cm]{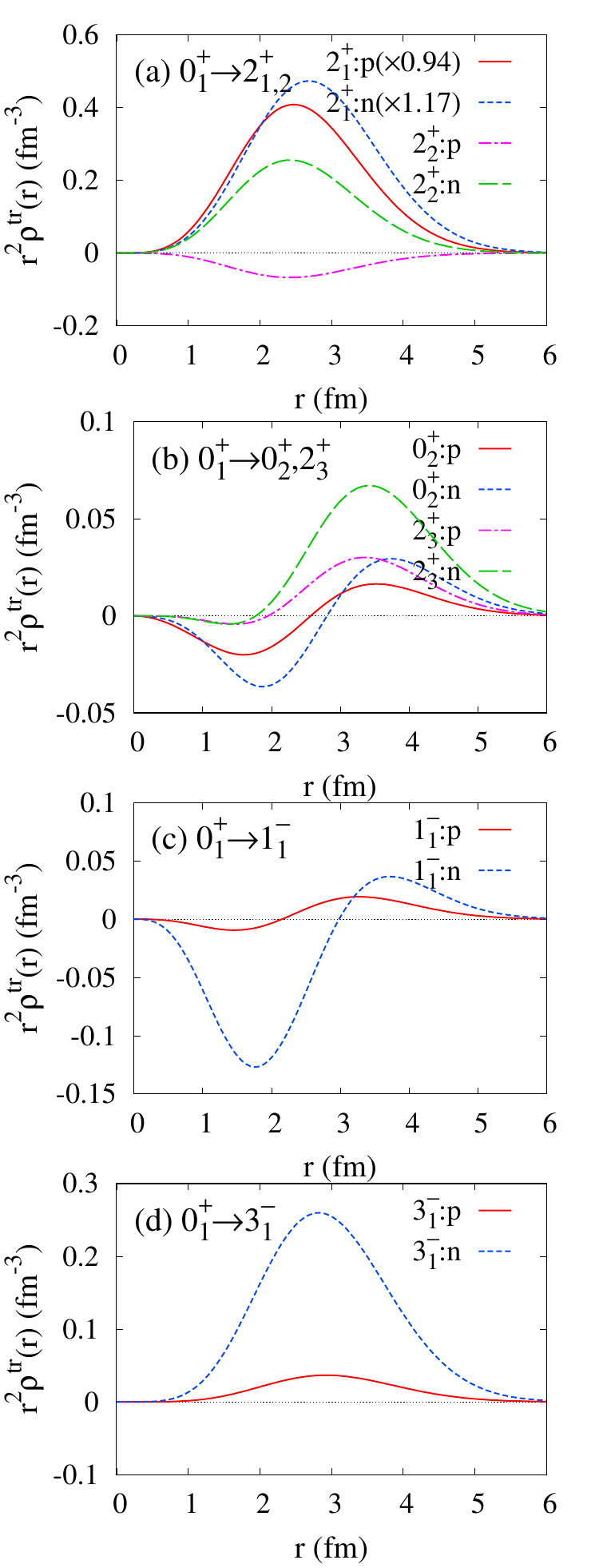}
  \caption{Transition densities of  $\Be$. The proton and neutron transition densities from the $0^+_1$ state
 to the (a) $2^+_{1,2}$, (b) $0^+_2$, $2^+_3$, (c) $1^-_1$, and (d) $3^-_1$ states. 
The $0^+_1\to 2^+_1$ transition densities are renormalized ones. 
  \label{fig:trans-be10}}
\end{figure}

The transition densities of $\Be$ are shown in Fig.~\ref{fig:trans-be10}.
Let us compare $2^+$ transitions from the $0^+_1$ state to 
the $2^+_1$($K^\pi=0^+_1$), $2^+_2$($K^\pi=2^+$), and $2^+_3$($K^\pi=0^+_2$) states. 
In the ground-band transition, $0^+_1\to 2^+_1$, the neutron transition density is similar to 
the proton one because this transition is the isoscalar excitation constructed 
by the $K=0$ rotation of the $2\alpha$ core part.
In other transitions, the amplitude of the neutron transition density is 
more than twice larger than that of the proton one showing the neutron dominance 
in the $2^+_2$ and $2^+_3$ excitations. 
Absolute amplitude of the neutron transition density is strongest in the ground-band $0^+_1\to 2^+_1$ transition, 
smaller in $0^+_1\to 2^+_2$, and further smaller in $0^+_1\to 2^+_3$.
One of the striking features is that, 
in the side-band transition, $0^+_1\to 2^+_2$, the proton component is 
opposite (negative sign) to the neutron one and gives 
cancellation effect to the isoscalar component, while 
the proton and neutron components are coherent in the $2^+_1$ and $2^+_3$ excitations.
In the radial behavior of the neutron transition density,  
one can see that the $0^+_1\to 2^+_2$ transition has a peak amplitude
slightly shifted inward compared with $0^+_1\to 2^+_1$ but the difference 
is not so remarkable. On the other hand, 
the $0^+_1\to 2^+_3$ transition has amplitude shifted to the outer region. 

In other inelastic transitions to the $0^+_2$, $1^-_1$, and $3^-_1$ states, 
the neutron transition density is dominant while the proton transition density 
is relatively weak indicating the neutron dominance. It should be commented that 
the $0^+_1\to 0^+_2$ and $0^+_1\to 1^-_1$ transitions show 
nodal structures as expected from the usual behavior of monopole and dipole transitions.


\subsection{$\Be+p$ and $\Be+\alpha$ reactions}

Using the AMD densities of  $\Be$, 
we perform the MCC calculations of the $\Be+p$ and $\Be+\alpha$ reactions.
For the coupled channels, we adopt the $0^+_{1,2}$, $2^+_{1.2.3}$, $1^-_1$, and $3^-_1$
states with $\lambda=0$, 1, 2, and 3 transitions between them. 
The experimental excitation energies of  $\Be$ are used.
For the $2^+_{1} \to 0^+_1$ transition, 
the renormalized transition densities are used as explained previously.
One-step (DWBA) cross sections are also calculated for comparison.
We also calculate the $^{10}\textrm{C}+p$ and $^{10}\textrm{C}+\alpha$ reactions
assuming the mirror symmetry of diagonal and transition densities between the proton and neutron parts
in the $A=10$ systems.
Coulomb shifts of  excitation energies are omitted.

\begin{figure*}[!h]
\includegraphics[width=18. cm]{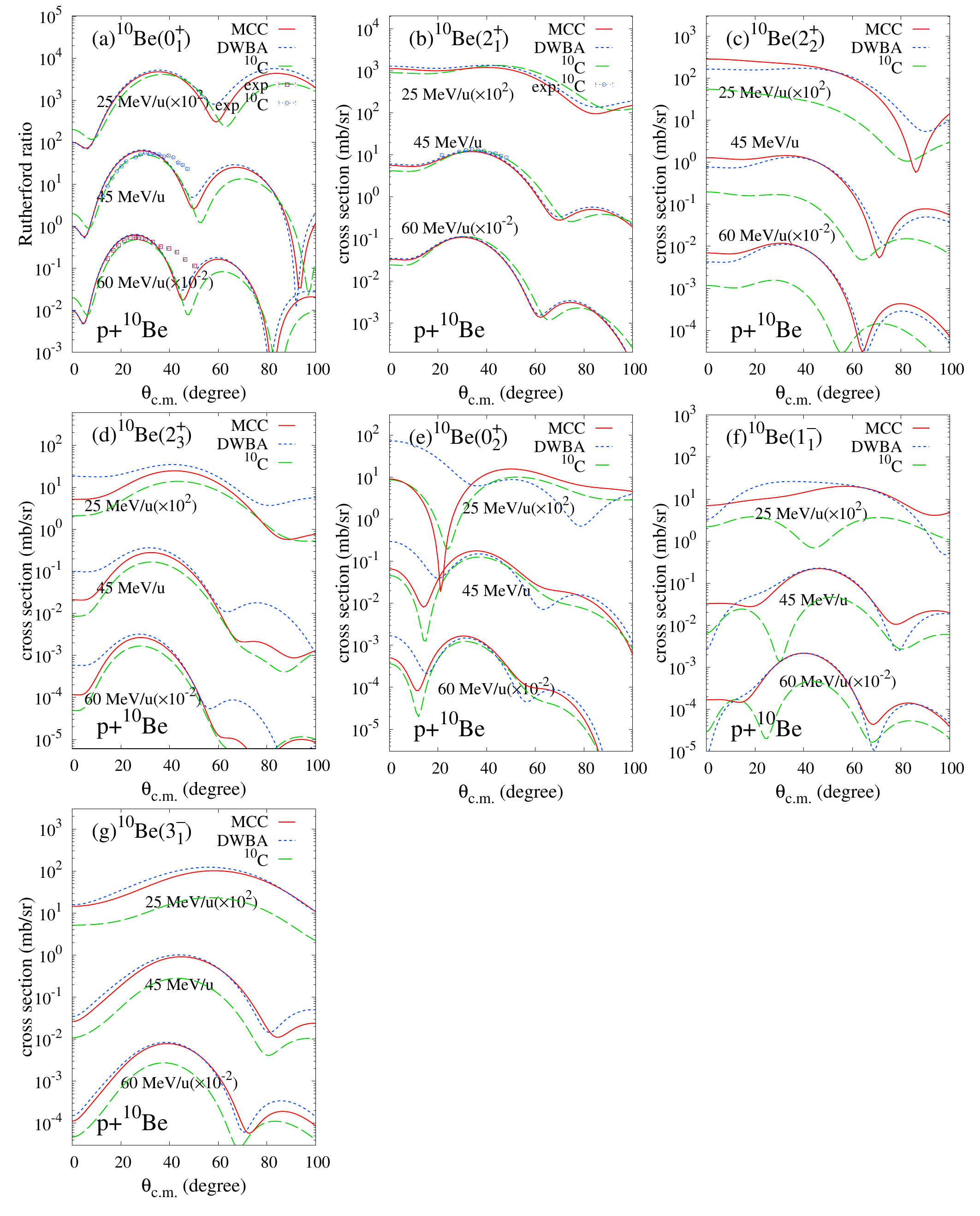}
  \caption{
Elastic and inelastic cross sections of the $\Be+p$ and $^{10}\textrm{C}+p$ reactions
 at $E=25$, 45, 60, and 100~MeV/u.
The MCC and DWBA results of $\Be+p$ are 
shown by red solid and blue dashed lines, respectively.
The MCC results of $^{10}\textrm{C}+p$ are shown by green long-dashed lines. 
The experimental $\Be+p$ elastic cross sections at $E=58.4$~MeV/u~\cite{CortinaGil:1997zk} are 
shown  by red squares in the panel (a).
The experimental $^{10}\textrm{C}+p$ cross sections at 
$E=45$~MeV/u~\cite{Jouanne:2005pb} observed for the $0^+_1$ and $2^+_1$ states are shown
 by blue circles in the panels (a) and (b).
  \label{fig:cross-be10p}}
\end{figure*}

\begin{figure*}[!h]
\includegraphics[width=18. cm]{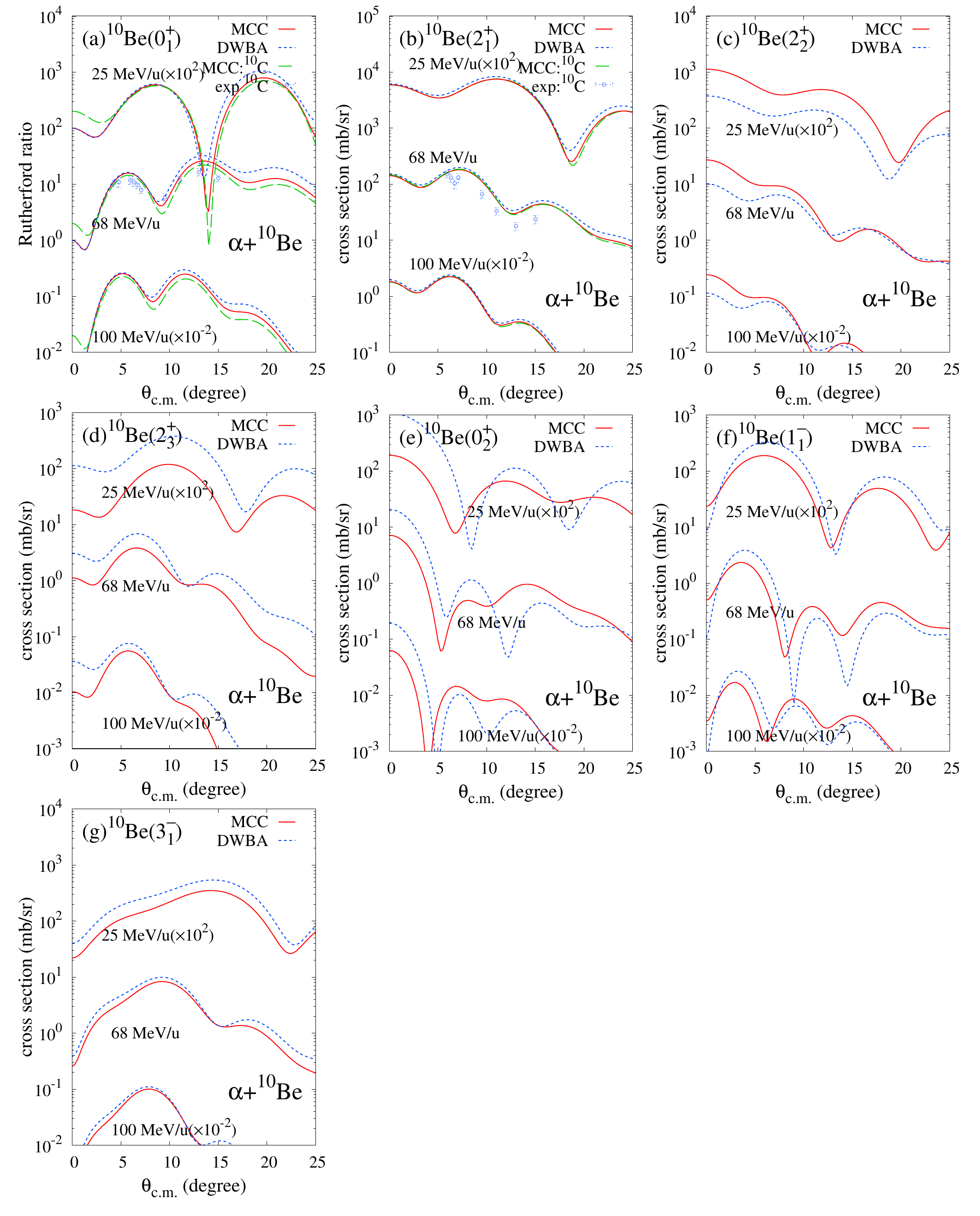}
  \caption{
Calculated elastic and inelastic cross sections of $\Be+\alpha$ at 
$E=25$, 68, and 100~MeV/u. 
The MCC and DWBA cross sections are 
shown by red solid and blue dashed lines, respectively. In the panels (a) and (b), the calculated 
cross sections of $^{10}\textrm{C}+\alpha$ are shown by green dashed lines compared with the 
experimental cross sections at $E=68$~MeV/u from Ref.~\cite{Furuno:2019lyp}. 
  \label{fig:cross-be10a}}
\end{figure*}

Figure~\ref{fig:cross-be10p} shows 
the calculated cross sections of  $\Be+p$ at $E=25$, 45, and 60~MeV/u 
together with those of $^{10}\textrm{C}+p$, and Fig.~\ref{fig:cross-be10a} shows the results of 
$\Be+\alpha$ at $E=25$, 68, and 100~MeV/u.
In Figs.~\ref{fig:cross-be10p}(a) and (b),  
the results are compared with the experimental $\Be+\alpha$ data of the elastic cross sections
at $E=60$~MeV/u~\cite{CortinaGil:1997zk} and the $^{10}\textrm{C}+p$ data of the elastic 
and $2^+_1$ cross sections at $E=45$~MeV/u~\cite{Jouanne:2005pb}, which have been 
observed by the inverse kinematics experiments.
The MCC calculations well reproduce those data 
as already shown in our previous work~\cite{Kanada-Enyo:2019uvg}. 
It should be noted again that the dip structure of 
elastic scattering can be smeared by the spin-orbit interaction omitted in the present calculation. 
In  Figs.~\ref{fig:cross-be10a}(a) and (b), we also show the result of the 
$^{10}\textrm{C}+\alpha$ reaction compared with the 
$^{10}\textrm{C}+\alpha$ data at $E=68$~MeV/u, which have been 
recently measured by the inverse kinematics experiment~\cite{Furuno:2019lyp}. 
The observed data of the elastic and $2^+_1$ cross sections tend to be smaller than the present result.
Comparing the MCC and DWBA results, one can see that 
CC effects are not minor except for the $2^+_1$ and $3^-_1$ cross sections of $\Be+p$ 
and the $2^+_1$ cross sections of $\Be+\alpha$. At low incident energies, 
remarkable CC effects can be seen in the $0^+_2$ cross sections of $\Be+p$ 
and the $2^+_2$, $2^+_3$, and $0^+_2$ cross sections of $\Be+\alpha$. The CC effects
enhance the $2^+_2$ cross sections and suppress the $0^+_2$ and $2^+_3$ cross sections. 
At higher incident energies, the CC effects become
weaker but they remain to be significant at forward angles even at
$E=60$~MeV/u of $\Be+p$ and $E=100$~MeV/u of $\Be+\alpha$. 

Let us compare $\Be+p$ and $^{10}\textrm{C}+p$ cross sections.
If a transition has the isoscalar character, 
difference between $\Be+p$ and  $^{10}\textrm{C}+p$ 
cross sections should be small. On the other hand, 
in the neutron dominant case, it is naively expected that
$\Be+p$ cross sections are enhanced and 
 $^{10}\textrm{C}+p$ cross sections are relatively suppressed 
because the $p$ scattering sensitively probes the neutron component 
rather than the proton component. 
In Fig.~\ref{fig:cross-be10p}, 
the $^{10}\textrm{C}+p$ cross sections (green dashed lines)  are 
compared with the  $\Be+p$ cross sections (red solid lines). As expected, 
the difference is small in the $2^+_1$ cross sections, 
because of the isoscalar nature of the ground-band transition. 
On the other hand, 
for the $2^+_2$, $1^-_1$, and $3^-_1$ states, 
the $^{10}\textrm{C}+p$ cross sections are 
remarkably suppressed 
compared with the $\Be+p$ cross sections
because of the neutron dominant characters of these transitions in $\Be$
(the proton dominance in $^{10}\textrm{C}$).

\begin{figure*}[!h]
\includegraphics[width=18 cm]{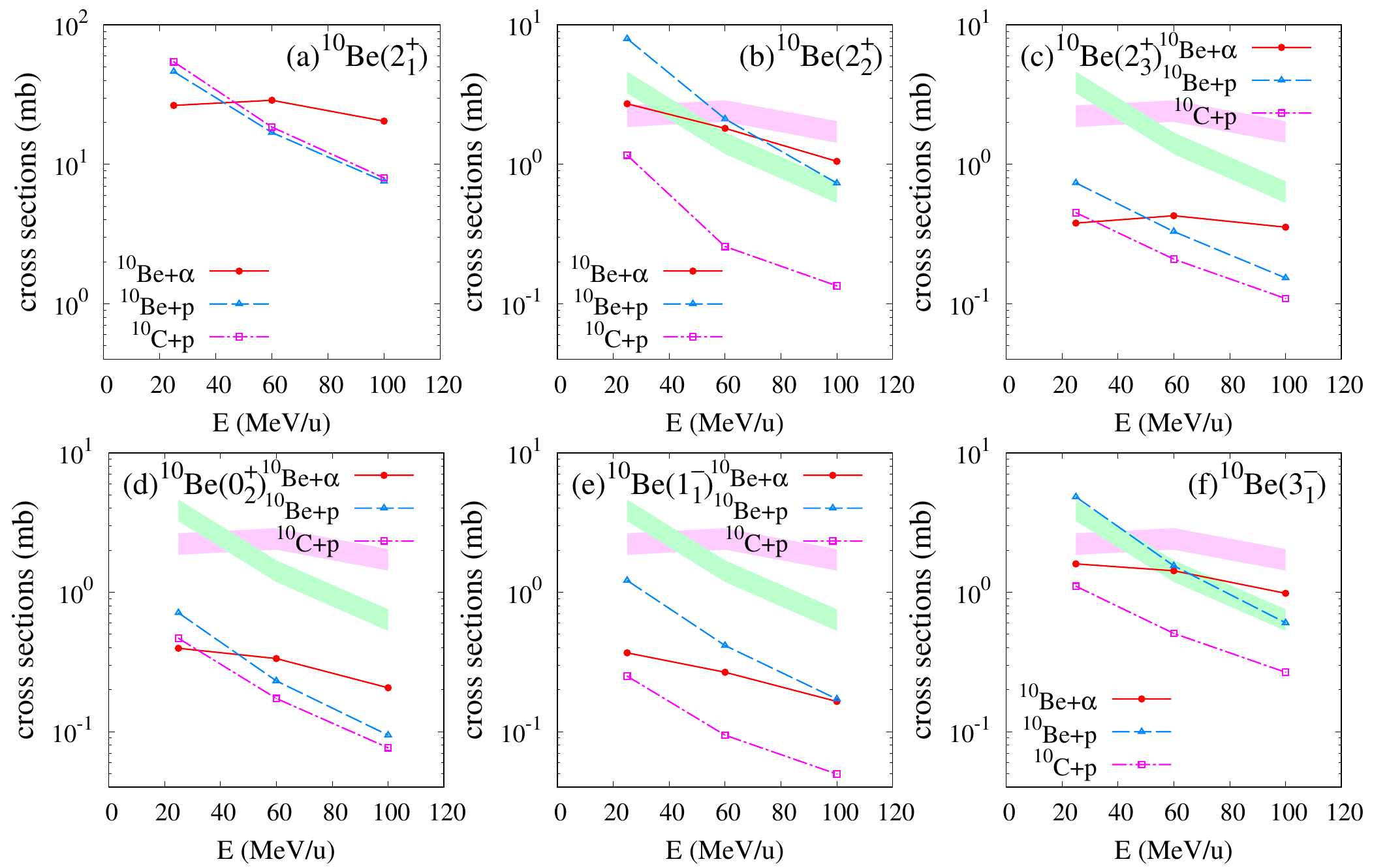}
  \caption{
Integrated cross sections of $\Be+p$, $^{10}\textrm{C}+p$, and $\Be+\alpha$ inelastic processes
are shown by (blue) triangles, (magenta) squares, and (red) circles, respectively. The cross sections at $E=25$, 60, and 100~MeV/u are calculated 
with MCC. 
In panels for the (b)~$2^+_2$, (c)~$2^+_3$, (d)~$0^+_2$, (e)~$1^-_1$, and (f)~$3^-_1$ states, 
7\%--10\% of the $2^+_1$ cross sections of the $\Be+p$ and  $\Be+\alpha$ reactions 
are shown by light-green and pink shaded areas, respectively, as references.
  \label{fig:cross-int-be10}}
\end{figure*}

For quantitative discussions, we compare the integrated 
cross sections of the inelastic scattering of the $\Be+p$, $^{10}\textrm{C}+p$ and $\Be+\alpha$
reactions. 
Figure~\ref{fig:cross-int-be10} shows the MCC results of the 
 cross sections at $E=25,$ 60, and 100~MeV/u.
For the $2^+_1$ cross sections, one can see only a small difference
between $\Be+p$ and $^{10}\textrm{C}+p$. 
This is a typical example of the isoscalar excitation and can be regarded as
reference data to be compared with other excitations. 
For the side-band $2^+_2$ sate,  
the difference between $\Be+p$ and $^{10}\textrm{C}+p$ is huge 
as one order of the magnitude of the cross sections 
because of the cancellation between proton and neutron components
in the $^{10}\textrm{C}+p$ reaction. 
As shown in the transition densities (Fig.~\ref{fig:trans-be10}(a))
and the matrix elements (Table~\ref{tab:BE2-be10}) of $\Be$, 
the proton component of the $2^+_2$ transition in $\Be$
is weak but opposite sign to the neutron one, and it gives the 
strong cancellation in the mirror transitions probed by the $^{10}\textrm{C}+p$ reaction. 
It also gives some cancellation in the isoscalar component probed by 
the $\Be+\alpha$ reaction, but the cancellation is tiny in the $\Be+p$ reaction. 
The difference of the production rates between the $\Be+p$ and $^{10}\textrm{C}+p$ reactions
is also large in the $3^-_1$ cross section
as expected from its remarkable neutron dominance (the ratio $M_n/M_p=6.4$).
Namely, the $3^-_1$ cross sections in the $\Be+p$ reaction 
are largely enhanced 
compared to the $^{10}\textrm{C}+p$ reaction.
Similarly, the enhancement of the $\Be+p$ cross sections 
is also obtained for the $0^+_2$ and $2^+_3$ states in the 
$K^\pi=0^+_2$ cluster-band, but it is not so remarkable as the $3^-_1$ state ($M_n/M_p=6.4$)
because of their weaker neutron dominance ($M_n/M_p=1.89$ of $0^+_1\to 0^+_2$ and 
$M_n/M_p=2.5$ of $0^+_1\to 2^+_3$). 
It is rather striking that,
the difference in the $1^-_1$ production rates between $\Be+p$ and $^{10}\textrm{C}+p$
is unexpectedly large even though the neutron dominance of the $1^-_1$ state
is weaker as $M_n/M_p=1.46$ than the $2^+_3$ and $0^+_2$ states. This is understood by 
the difference in radial behaviors of the proton and neutron transition densities. As 
shown in Fig.~\ref{fig:trans-be10}(c) for the $0^+_1\to 1^-_1$ transition densities, 
the neutron amplitude is dominant in the outer region and enhances the $\Be+p$ cross sections. 
Moreover, at the surface region of $r=2$--3 fm, 
the proton transition density is opposite to the neutron one and gives 
the cancellation effect in the $^{10}\textrm{C}+p$ reaction.

In the experimental side, 
the $(p,p')$ and $(\alpha,\alpha')$ cross sections off $\Be$ and $^{10}\textrm{C}$ 
have been measured only for the $2^+_1$ state. Indeed, according to the present calculation,  
the $2^+_1$ state is strongly populated in 
$p$ and $\alpha$ inelastic scattering processes,
but other states are relatively weak as more than one order 
smaller cross sections than the  $2^+_1$ state. Below, we 
discuss sensitivity of the $\Be+p$, $\Be+\alpha$, and $^{10}\textrm{C}+p$
reactions to observe higher excited states above the $2^+_1$ state. 

Firstly, we examine the integrated cross sections
and discuss the production rates of excited states and their projectile and energy dependencies. 
In Figs.~\ref{fig:cross-int-be10}(b)-(f), 
7\%--10\% of the $2^+_1$ cross sections of the $\Be+p$ and $\Be+\alpha$ reactions 
are shown by light-green and pink shaded areas, respectively. We consider 
these areas as references of one-order smaller magnitude of the $2^+_1$ cross sections
for comparison. 
For the side-band $2^+_2$ transition (Fig.~\ref{fig:cross-int-be10}(b)), 
the $\Be+p$ cross sections (blue dashed line) exceed the 7\%--10\% area (light-green)
indicating that the $\Be+p$ reaction can be an efficient tool to 
observe the neutron dominance of the $2^+_2$ excitation.
Also the $\Be+\alpha$ cross sections (red solid lines) 
reach 10\% of the $2^+_1$ cross sections at $E= 25$~MeV/u but 
decrease at high energies. 
For the $3^-_1$ transitions (Fig.~\ref{fig:cross-int-be10}(f)), the $\Be+p$ cross sections (blue dashed line) 
are within the 7\%--10\% area (light-green), and the $\Be+\alpha$ cross sections are 
approximately 5\% of the  $2^+_1$ cross sections. 
For other states, the population is much weaker as 1--2\% of the $2^+_1$ state or less.

\begin{figure}[!h]
\includegraphics[width=7 cm]{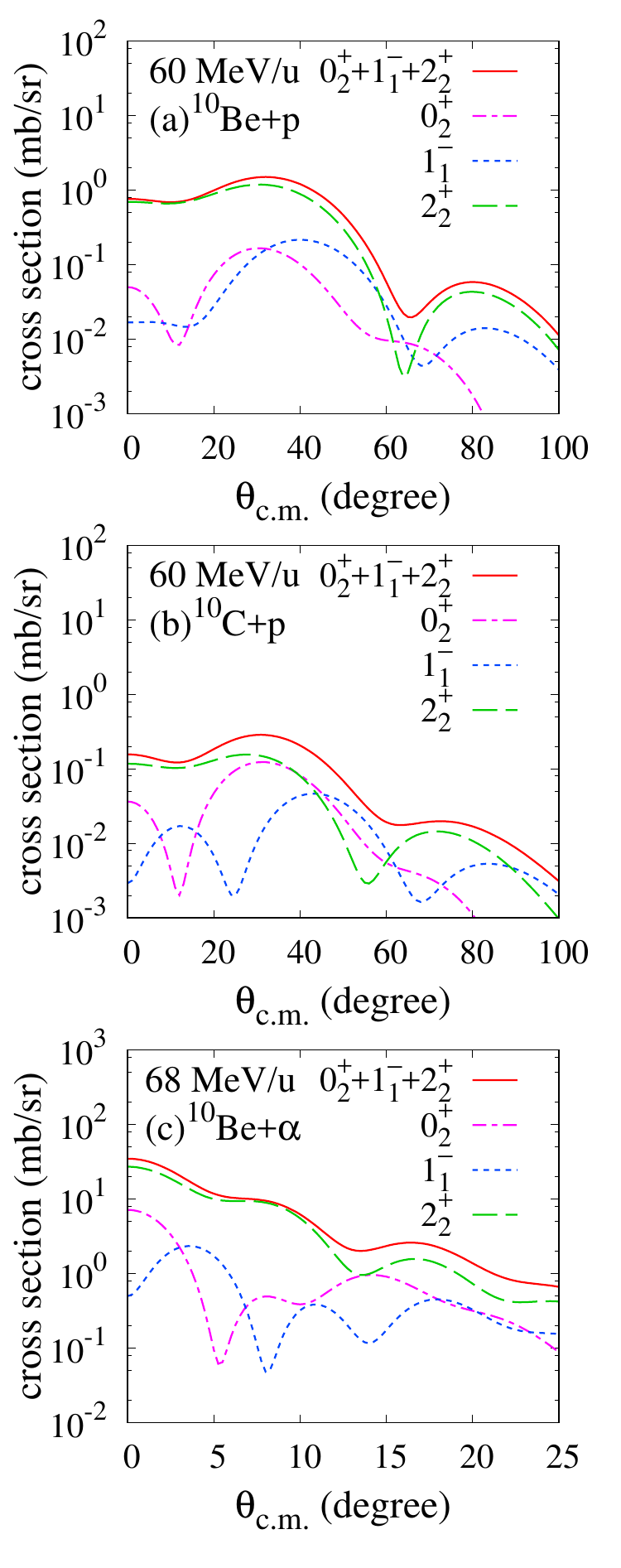}
  \caption{
The $0^+_2$, $1^-_1$, and $2^+_2$
cross sections of the $\Be+p$, $^{10}\textrm{C}+p$, and $\Be+\alpha$ reactions 
obtained by the MCC calculations. (a)$\Be+p$ at $E=60$~MeV/u, 
(b)$^{10}\textrm{C}+p$ at $E=60$~MeV/u, and (c)$\Be+\alpha$ at $E=68$~MeV/u. 
The $0^+_2$, $1^-_1$, and $2^+_2$ cross sections and their incoherent sum are 
shown by magenta dot-dashed, blue dashed, green long dashed, and red solid lines. 
  \label{fig:cross-be10a-012}}
\end{figure}

Next, we compare the $0^+_2$, $2^+_2$, and $1^-_1$ cross sections of each reaction. 
Since these three states almost degenerate around $E_x\approx 6$ MeV
in the experimental energy spectra,  
it may be difficult to resolve observed cross sections to individual states.
In Figs.~\ref{fig:cross-be10a-012}(a), (b), and (c), the calculated cross sections of 
$\Be+p$ at $E$=60~MeV/u,  $^{10}\textrm{C}+p$ at  $E$=60~MeV/u,
and $\Be+\alpha$ at $E$=68~MeV/u are shown, respectively. 
The cross sections of each state and the incoherent sum of three states 
are plotted.
In the $\Be+p$ reaction, the $2^+_2$ cross sections dominate the summed cross sections while
the $0^+_2$ and $1^-_1$ contributions are minor. 
In the $^{10}\textrm{C}+p$ reaction, where the  $2^+_2$ cross sections are strongly suppressed, 
the magnitude of the $0^+_2$ cross sections is comparable to that of $2^+_2$ 
in the $\theta_\textrm{c.m.}=20$--$40^\circ$ range, 
and the $1^-_1$ state gives major contribution at $\theta_\textrm{c.m.}\sim 50^\circ$
and smears the second dip of the $2^+_2$ cross sections in the summed cross sections.
In the $\Be+\alpha$ reaction at forward angles,  
the $0^+_2$ and $1^-_1$ contributions are minor compared to the dominant 
$2^+_2$ contribution. 
It seems to contradict the usual expectation that
forward angle $\alpha$ scattering can be generally useful  to observe monopole transitions.
But it is not the case in the $\Be+\alpha$ reaction because 
the $0^+_2$ cross sections at 
forward angles are strongly suppressed by the CC effect. 
Alternatively, 
detailed analysis of $^{10}\textrm{C}+p$ cross sections in a wide range of scattering angles
may be promising to observe the $0^+_2$ and $1^-_1$ states. 

It should be commented that the predicted cross sections still contain structure model 
ambiguity, in particular, for the cluster-bands. 
Basis configurations adopted in the present AMD calculation 
are not enough to describe 
details of the inter-cluster motion, 
which may somewhat enhance the monopole transition strengths. 

\section{Summary} \label{sec:summary}
Isospin characters of nuclear excitations in $^{26}$Mg and $^{10}$Be were investigated with 
the MCC calculations of $p$ and $\alpha$ inelastic scattering. 
The structure calculations of $^{26}$Mg and $^{10}$Be were done by 
antisymmetrized molecular dynamics (AMD).
In the AMD calculations, the $K^\pi=0^+$ ground- and $K^\pi=2^+$ side-bands were obtained in $^{26}$Mg and $^{10}$Be.
In both systems, 
the ground-band $2^+_1$($K^\pi=0^+_1$)  state and
 the side-band $2^+_2$($K^\pi=2^+$) state have 
quite different isospin characters.
The former has the isoscalar feature and the latter shows the neutron dominance character.
This can be a general feature in $N=Z+2$ system having a prolately deformed 
$N=Z$ core  surrounded by valence neutrons. 

The MCC calculations of $p$ and $\alpha$ inelastic scattering off $^{26}$Mg and $^{10}$Be 
were performed with the Melbourne $g$-matrix folding approach 
by using the matter and transition densities of the target nuclei calculated with AMD.
The calculations reasonably reproduced
the observed $0^+_1$, $2^+_1$, and $2^+_2$ cross sections of $^{26}$Mg+$p$ scattering 
at $E_p=24$ and 40~MeV and of $^{26}$Mg+$\alpha$ scattering at $E_\alpha=104$ and 120~MeV.
It was shown that the $^{26}$Mg+$p$ scattering is a sensitive probe to the neutron component of 
the $0^+_1\to 2^+_2$ transition.
In the present analysis, the neutron transition matrix element $M_n\sim 8$~fm$^2$ 
(the squared ratio $|M_n/M_p|^2\sim 7$) of the $0^+_1\to 2^+_2$ transitions in $\Mg$ 
is favored to reproduce the $\Mg+p$ and  $\Mg+\alpha$ cross sections consistently. 

For $^{10}$Be+$p$ and  $^{10}$Be+$\alpha$ scattering, inelastic cross sections to the excited states in 
the $K^\pi=0^+_1$ ground-, $K^\pi=2^+$ side-, 
$K^\pi=0^+_2$ cluster-, and $K^\pi=1^-$ cluster-bands were discussed.
In a comparison of the $\Be+p$, $^{10}\textrm{C}+p$, and $\Be+\alpha$ reactions, 
the isospin characters of transitions in inelastic scattering processes were investigated. 
Also in $\Be$, 
the $p$ inelastic scattering was found to be a sensitive probe to the neutron dominance in 
the $2^+_2$ excitation. 
The significant suppression of the $2^+_2$ cross sections of $^{10}\textrm{C}+p$ was obtained 
because of the cancellation of the proton and neutron components in the transition.
The present prediction of the inelastic scattering off $\Be$ may
be useful for the feasibility test of future experiments in the inverse kinematics.

\begin{acknowledgments}
The authors thank Dr.~Furuno and Dr.~Kawabata for fruitful discussions.
The computational calculations of this work were performed by using the
supercomputer in the Yukawa Institute for theoretical physics, Kyoto University. This work was partly supported
by Grants-in-Aid of the Japan Society for the Promotion of Science (Grant Nos. JP18K03617, JP16K05352, and 18H05407) and by the grant for the RCNP joint research project.
\end{acknowledgments}

\end{document}